\documentclass[11pt]{article}
\usepackage{amsmath}
\usepackage{amssymb}

\def\PsfigVersion{1.9}
\ifx\undefined\psfig\else\endinput\fi

\let\LaTeXAtSign=\@
\let\@=\relax
\edef\psfigRestoreAt{\catcode`\@=\number\catcode`@\relax}
\catcode`\@=11\relax
\newwrite\@unused
\def\ps@typeout#1{{\let\protect\string\immediate\write\@unused{#1}}}
\ps@typeout{psfig/tex \PsfigVersion}

\def\figurepath{./}

\def\@nnil{\@nil}
\def\@empty{}
\def\@psdonoop#1\@@#2#3{}
\def\@psdo#1:=#2\do#3{\edef\@psdotmp{#2}\ifx\@psdotmp\@empty \else
    \expandafter\@psdoloop#2,\@nil,\@nil\@@#1{#3}\fi}
\def\@psdoloop#1,#2,#3\@@#4#5{\def#4{#1}\ifx #4\@nnil \else
       #5\def#4{#2}\ifx #4\@nnil \else#5\@ipsdoloop #3\@@#4{#5}\fi\fi}
\def\@ipsdoloop#1,#2\@@#3#4{\def#3{#1}\ifx #3\@nnil 
       \let\@nextwhile=\@psdonoop \else
      #4\relax\let\@nextwhile=\@ipsdoloop\fi\@nextwhile#2\@@#3{#4}}
\def\@tpsdo#1:=#2\do#3{\xdef\@psdotmp{#2}\ifx\@psdotmp\@empty \else
    \@tpsdoloop#2\@nil\@nil\@@#1{#3}\fi}
\def\@tpsdoloop#1#2\@@#3#4{\def#3{#1}\ifx #3\@nnil 
       \let\@nextwhile=\@psdonoop \else
      #4\relax\let\@nextwhile=\@tpsdoloop\fi\@nextwhile#2\@@#3{#4}}
\ifx\undefined\fbox
\newdimen\fboxrule
\newdimen\fboxsep
\newdimen\ps@tempdima
\newbox\ps@tempboxa
\long\def\fbox#1{\leavevmode\setbox\ps@tempboxa\hbox{#1}\ps@tempdima\fboxrule
    \advance\ps@tempdima \fboxsep \advance\ps@tempdima \dp\ps@tempboxa
   \hbox{\lower \ps@tempdima\hbox
  {\vbox{\hrule height \fboxrule
          \hbox{\vrule width \fboxrule \hskip\fboxsep
          \vbox{\vskip\fboxsep \box\ps@tempboxa\vskip\fboxsep}\hskip 
                 \fboxsep\vrule width \fboxrule}
                 \hrule height \fboxrule}}}}
\fi
\newread\ps@stream
\newif\ifnot@eof       
\newif\if@noisy        
\newif\if@atend        
\newif\if@psfile       
{\catcode`\%=12\global\gdef\epsf@start{
\def\epsf@PS{PS}
\def\epsf@getbb#1{%
\openin\ps@stream=#1
\ifeof\ps@stream\ps@typeout{Error, File #1 not found}\else
   {\not@eoftrue \chardef\other=12
    \def\do##1{\catcode`##1=\other}\dospecials \catcode`\ =10
    \loop
       \if@psfile
          \read\ps@stream to \epsf@fileline
       \else{
          \obeyspaces
          \read\ps@stream to \epsf@tmp\global\let\epsf@fileline\epsf@tmp}
       \fi
       \ifeof\ps@stream\not@eoffalse\else
       \if@psfile\else
       \expandafter\epsf@test\epsf@fileline:. \\%
       \fi
          \expandafter\epsf@aux\epsf@fileline:. \\%
       \fi
   \ifnot@eof\repeat
   }\closein\ps@stream\fi}%
\long\def\epsf@test#1#2#3:#4\\{\def\epsf@testit{#1#2}
                        \ifx\epsf@testit\epsf@start\else
\ps@typeout{Warning! File does not start with `\epsf@start'.  It may not be a PostScript file.}
                        \fi
                        \@psfiletrue} 
{\catcode`\%=12\global\let\epsf@percent=
\long\def\epsf@aux#1#2:#3\\{\ifx#1\epsf@percent
   \def\epsf@testit{#2}\ifx\epsf@testit\epsf@bblit
        \@atendfalse
        \epsf@atend #3 . \\%
        \if@atend       
           \if@verbose{
                \ps@typeout{psfig: found `(atend)'; continuing search}
           }\fi
        \else
        \epsf@grab #3 . . . \\%
        \not@eoffalse
        \global\no@bbfalse
        \fi
   \fi\fi}%
\def\epsf@grab #1 #2 #3 #4 #5\\{%
   \global\def\epsf@llx{#1}\ifx\epsf@llx\empty
      \epsf@grab #2 #3 #4 #5 .\\\else
   \global\def\epsf@lly{#2}%
   \global\def\epsf@urx{#3}\global\def\epsf@ury{#4}\fi}%
\def\epsf@atendlit{(atend)} 
\def\epsf@atend #1 #2 #3\\{%
   \def\epsf@tmp{#1}\ifx\epsf@tmp\empty
      \epsf@atend #2 #3 .\\\else
   \ifx\epsf@tmp\epsf@atendlit\@atendtrue\fi\fi}

\chardef\psletter = 11 
\chardef\other = 12

\newif \ifdebug 
\newif\ifc@mpute 
\c@mputetrue 

\let\then = \relax
\def\r@dian{pt }
\let\r@dians = \r@dian
\let\dimensionless@nit = \r@dian
\let\dimensionless@nits = \dimensionless@nit
\def\internal@nit{sp }
\let\internal@nits = \internal@nit
\newif\ifstillc@nverging
\def \Mess@ge #1{\ifdebug \then \message {#1} \fi}

{ 
        \catcode `\@ = \psletter
        \gdef \nodimen {\expandafter \n@dimen \the \dimen}
        \gdef \term #1 #2 #3%
               {\edef \t@ {\the #1}
                \edef \t@@ {\expandafter \n@dimen \the #2\r@dian}%
                \t@rm {\t@} {\t@@} {#3}%
               }
        \gdef \t@rm #1 #2 #3%
               {{%
                \count 0 = 0
                \dimen 0 = 1 \dimensionless@nit
                \dimen 2 = #2\relax
                \Mess@ge {Calculating term #1 of \nodimen 2}%
                \loop
                \ifnum  \count 0 < #1
                \then   \advance \count 0 by 1
                        \Mess@ge {Iteration \the \count 0 \space}%
                        \Multiply \dimen 0 by {\dimen 2}%
                        \Mess@ge {After multiplication, term = \nodimen 0}%
                        \Divide \dimen 0 by {\count 0}%
                        \Mess@ge {After division, term = \nodimen 0}%
                \repeat
                \Mess@ge {Final value for term #1 of 
                                \nodimen 2 \space is \nodimen 0}%
                \xdef \Term {#3 = \nodimen 0 \r@dians}%
                \aftergroup \Term
               }}
        \catcode `\p = \other
        \catcode `\t = \other
        \gdef \n@dimen #1pt{#1} 
}

\def \Divide #1by #2{\divide #1 by #2} 

\def \Multiply #1by #2
       {{
        \count 0 = #1\relax
        \count 2 = #2\relax
        \count 4 = 65536
        \Mess@ge {Before scaling, count 0 = \the \count 0 \space and
                        count 2 = \the \count 2}%
        \ifnum  \count 0 > 32767 
        \then   \divide \count 0 by 4
                \divide \count 4 by 4
        \else   \ifnum  \count 0 < -32767
                \then   \divide \count 0 by 4
                        \divide \count 4 by 4
                \else
                \fi
        \fi
        \ifnum  \count 2 > 32767 
        \then   \divide \count 2 by 4
                \divide \count 4 by 4
        \else   \ifnum  \count 2 < -32767
                \then   \divide \count 2 by 4
                        \divide \count 4 by 4
                \else
                \fi
        \fi
        \multiply \count 0 by \count 2
        \divide \count 0 by \count 4
        \xdef \product {#1 = \the \count 0 \internal@nits}%
        \aftergroup \product
       }}

\def\r@duce{\ifdim\dimen0 > 90\r@dian \then   
                \multiply\dimen0 by -1
                \advance\dimen0 by 180\r@dian
                \r@duce
            \else \ifdim\dimen0 < -90\r@dian \then  
                \advance\dimen0 by 360\r@dian
                \r@duce
                \fi
            \fi}

\def\Sine#1%
       {{%
        \dimen 0 = #1 \r@dian
        \r@duce
        \ifdim\dimen0 = -90\r@dian \then
           \dimen4 = -1\r@dian
           \c@mputefalse
        \fi
        \ifdim\dimen0 = 90\r@dian \then
           \dimen4 = 1\r@dian
           \c@mputefalse
        \fi
        \ifdim\dimen0 = 0\r@dian \then
           \dimen4 = 0\r@dian
           \c@mputefalse
        \fi
        \ifc@mpute \then
                \divide\dimen0 by 180
                \dimen0=3.141592654\dimen0
                \dimen 2 = 3.1415926535897963\r@dian 
                \divide\dimen 2 by 2 
                \Mess@ge {Sin: calculating Sin of \nodimen 0}%
                \count 0 = 1 
                \dimen 2 = 1 \r@dian 
                \dimen 4 = 0 \r@dian 
                \loop
                        \ifnum  \dimen 2 = 0 
                        \then   \stillc@nvergingfalse 
                        \else   \stillc@nvergingtrue
                        \fi
                        \ifstillc@nverging 
                        \then   \term {\count 0} {\dimen 0} {\dimen 2}%
                                \advance \count 0 by 2
                                \count 2 = \count 0
                                \divide \count 2 by 2
                                \ifodd  \count 2 
                                \then   \advance \dimen 4 by \dimen 2
                                \else   \advance \dimen 4 by -\dimen 2
                                \fi
                \repeat
        \fi             
                        \xdef \sine {\nodimen 4}%
       }}

\def\Cosine#1{\ifx\sine\UnDefined\edef\Savesine{\relax}\else
                             \edef\Savesine{\sine}\fi
        {\dimen0=#1\r@dian\advance\dimen0 by 90\r@dian
         \Sine{\nodimen 0}
         \xdef\cosine{\sine}
         \xdef\sine{\Savesine}}}              

\def\psdraft{
        \def\@psdraft{0}
}
\def\psfull{
        \def\@psdraft{100}
}

\psfull

\newif\if@scalefirst
\def\psscalefirst{\@scalefirsttrue}
\def\psrotatefirst{\@scalefirstfalse}
\psrotatefirst

\newif\if@draftbox
\def\psnodraftbox{
        \@draftboxfalse
}
\def\psdraftbox{
        \@draftboxtrue
}
\@draftboxtrue

\newif\if@prologfile
\newif\if@postlogfile
\def\pssilent{
        \@noisyfalse
}
\def\psnoisy{
        \@noisytrue
}
\psnoisy
\newif\if@bbllx
\newif\if@bblly
\newif\if@bburx
\newif\if@bbury
\newif\if@height
\newif\if@width
\newif\if@rheight
\newif\if@rwidth
\newif\if@angle
\newif\if@clip
\newif\if@verbose
\def\@p@@sclip#1{\@cliptrue}

\newif\if@decmpr

\def\@p@@sfigure#1{\def\@p@sfile{null}\def\@p@sbbfile{null}
                \openin1=#1.bb
                \ifeof1\closein1
                        \openin1=\figurepath#1.bb
                        \ifeof1\closein1
                                \openin1=#1
                                \ifeof1\closein1%
                                       \openin1=\figurepath#1
                                        \ifeof1
                                           \ps@typeout{Error, File #1 not found}
                                                \if@bbllx\if@bblly
                                                \if@bburx\if@bbury
                                                        \def\@p@sfile{#1}%
                                                        \def\@p@sbbfile{#1}%
                                                        \@decmprfalse
                                                \fi\fi\fi\fi
                                        \else\closein1
                                                \def\@p@sfile{\figurepath#1}%
                                                \def\@p@sbbfile{\figurepath#1}%
                                                \@decmprfalse
                                        \fi%
                                \else\closein1%
                                        \def\@p@sfile{#1}
                                        \def\@p@sbbfile{#1}
                                        \@decmprfalse
                                \fi
                        \else
                                \def\@p@sfile{\figurepath#1}
                                \def\@p@sbbfile{\figurepath#1.bb}
                                \@decmprtrue
                        \fi
                \else
                        \def\@p@sfile{#1}
                        \def\@p@sbbfile{#1.bb}
                        \@decmprtrue
                \fi}

\def\@p@@sfile#1{\@p@@sfigure{#1}}

\def\@p@@sbbllx#1{
                \@bbllxtrue
                \dimen100=#1
                \edef\@p@sbbllx{\number\dimen100}
}
\def\@p@@sbblly#1{
                \@bbllytrue
                \dimen100=#1
                \edef\@p@sbblly{\number\dimen100}
}
\def\@p@@sbburx#1{
                \@bburxtrue
                \dimen100=#1
                \edef\@p@sbburx{\number\dimen100}
}
\def\@p@@sbbury#1{
                \@bburytrue
                \dimen100=#1
                \edef\@p@sbbury{\number\dimen100}
}
\def\@p@@sheight#1{
                \@heighttrue
                \dimen100=#1
                \edef\@p@sheight{\number\dimen100}
}
\def\@p@@swidth#1{
                \@widthtrue
                \dimen100=#1
                \edef\@p@swidth{\number\dimen100}
}
\def\@p@@srheight#1{
                \@rheighttrue
                \dimen100=#1
                \edef\@p@srheight{\number\dimen100}
}
\def\@p@@srwidth#1{
                \@rwidthtrue
                \dimen100=#1
                \edef\@p@srwidth{\number\dimen100}
}
\def\@p@@sangle#1{
                \@angletrue
                \edef\@p@sangle{#1} 
}
\def\@p@@ssilent#1{ 
                \@verbosefalse
}
\def\@p@@sprolog#1{\@prologfiletrue\def\@prologfileval{#1}}
\def\@p@@spostlog#1{\@postlogfiletrue\def\@postlogfileval{#1}}
\def\@cs@name#1{\csname #1\endcsname}
\def\@setparms#1=#2,{\@cs@name{@p@@s#1}{#2}}
\def\ps@init@parms{
                \@bbllxfalse \@bbllyfalse
                \@bburxfalse \@bburyfalse
                \@heightfalse \@widthfalse
                \@rheightfalse \@rwidthfalse
                \def\@p@sbbllx{}\def\@p@sbblly{}
                \def\@p@sbburx{}\def\@p@sbbury{}
                \def\@p@sheight{}\def\@p@swidth{}
                \def\@p@srheight{}\def\@p@srwidth{}
                \def\@p@sangle{0}
                \def\@p@sfile{} \def\@p@sbbfile{}
                \def\@p@scost{10}
                \def\@sc{}
                \@prologfilefalse
                \@postlogfilefalse
                \@clipfalse
                \if@noisy
                        \@verbosetrue
                \else
                        \@verbosefalse
                \fi
}
\def\parse@ps@parms#1{
                \@psdo\@psfiga:=#1\do
                   {\expandafter\@setparms\@psfiga,}}
\newif\ifno@bb
\def\bb@missing{
        \if@verbose{
                \ps@typeout{psfig: searching \@p@sbbfile \space  for bounding box}
        }\fi
        \no@bbtrue
        \epsf@getbb{\@p@sbbfile}
        \ifno@bb \else \bb@cull\epsf@llx\epsf@lly\epsf@urx\epsf@ury\fi
}       
\def\bb@cull#1#2#3#4{
        \dimen100=#1 bp\edef\@p@sbbllx{\number\dimen100}
        \dimen100=#2 bp\edef\@p@sbblly{\number\dimen100}
        \dimen100=#3 bp\edef\@p@sbburx{\number\dimen100}
        \dimen100=#4 bp\edef\@p@sbbury{\number\dimen100}
        \no@bbfalse
}
\newdimen\p@intvaluex
\newdimen\p@intvaluey
\def\rotate@#1#2{{\dimen0=#1 sp\dimen1=#2 sp
                  \global\p@intvaluex=\cosine\dimen0
                  \dimen3=\sine\dimen1
                  \global\advance\p@intvaluex by -\dimen3
                  \global\p@intvaluey=\sine\dimen0
                  \dimen3=\cosine\dimen1
                  \global\advance\p@intvaluey by \dimen3
                  }}
\def\compute@bb{
                \no@bbfalse
                \if@bbllx \else \no@bbtrue \fi
                \if@bblly \else \no@bbtrue \fi
                \if@bburx \else \no@bbtrue \fi
                \if@bbury \else \no@bbtrue \fi
                \ifno@bb \bb@missing \fi
                \ifno@bb \ps@typeout{FATAL ERROR: no bb supplied or found}
                        \no-bb-error
                \fi
                \count203=\@p@sbburx
                \count204=\@p@sbbury
                \advance\count203 by -\@p@sbbllx
                \advance\count204 by -\@p@sbblly
                \edef\ps@bbw{\number\count203}
                \edef\ps@bbh{\number\count204}
                \if@angle 
                        \Sine{\@p@sangle}\Cosine{\@p@sangle}
                        {\dimen100=\maxdimen\xdef\r@p@sbbllx{\number\dimen100}
                                            \xdef\r@p@sbblly{\number\dimen100}
                                            \xdef\r@p@sbburx{-\number\dimen100}
                                            \xdef\r@p@sbbury{-\number\dimen100}}
                        \def\minmaxtest{
                           \ifnum\number\p@intvaluex<\r@p@sbbllx
                              \xdef\r@p@sbbllx{\number\p@intvaluex}\fi
                           \ifnum\number\p@intvaluex>\r@p@sbburx
                              \xdef\r@p@sbburx{\number\p@intvaluex}\fi
                           \ifnum\number\p@intvaluey<\r@p@sbblly
                              \xdef\r@p@sbblly{\number\p@intvaluey}\fi
                           \ifnum\number\p@intvaluey>\r@p@sbbury
                              \xdef\r@p@sbbury{\number\p@intvaluey}\fi
                           }
                        \rotate@{\@p@sbbllx}{\@p@sbblly}
                        \minmaxtest
                        \rotate@{\@p@sbbllx}{\@p@sbbury}
                        \minmaxtest
                        \rotate@{\@p@sbburx}{\@p@sbblly}
                        \minmaxtest
                        \rotate@{\@p@sbburx}{\@p@sbbury}
                        \minmaxtest
                        \edef\@p@sbbllx{\r@p@sbbllx}\edef\@p@sbblly{\r@p@sbblly}
                        \edef\@p@sbburx{\r@p@sbburx}\edef\@p@sbbury{\r@p@sbbury}
                \fi
                \count203=\@p@sbburx
                \count204=\@p@sbbury
                \advance\count203 by -\@p@sbbllx
                \advance\count204 by -\@p@sbblly
                \edef\@bbw{\number\count203}
                \edef\@bbh{\number\count204}
}
\def\in@hundreds#1#2#3{\count240=#2 \count241=#3
                     \count100=\count240        
                     \divide\count100 by \count241
                     \count101=\count100
                     \multiply\count101 by \count241
                     \advance\count240 by -\count101
                     \multiply\count240 by 10
                     \count101=\count240        
                     \divide\count101 by \count241
                     \count102=\count101
                     \multiply\count102 by \count241
                     \advance\count240 by -\count102
                     \multiply\count240 by 10
                     \count102=\count240        
                     \divide\count102 by \count241
                     \count200=#1\count205=0
                     \count201=\count200
                        \multiply\count201 by \count100
                        \advance\count205 by \count201
                     \count201=\count200
                        \divide\count201 by 10
                        \multiply\count201 by \count101
                        \advance\count205 by \count201
                     \count201=\count200
                        \divide\count201 by 100
                        \multiply\count201 by \count102
                        \advance\count205 by \count201
                     \edef\@result{\number\count205}
}
\def\compute@wfromh{
                \in@hundreds{\@p@sheight}{\@bbw}{\@bbh}
                \edef\@p@swidth{\@result}
}
\def\compute@hfromw{
                \in@hundreds{\@p@swidth}{\@bbh}{\@bbw}
                \edef\@p@sheight{\@result}
}
\def\compute@handw{
                \if@height 
                        \if@width
                        \else
                                \compute@wfromh
                        \fi
                \else 
                        \if@width
                                \compute@hfromw
                        \else
                                \edef\@p@sheight{\@bbh}
                                \edef\@p@swidth{\@bbw}
                        \fi
                \fi
}
\def\compute@resv{
                \if@rheight \else \edef\@p@srheight{\@p@sheight} \fi
                \if@rwidth \else \edef\@p@srwidth{\@p@swidth} \fi
}
\def\compute@sizes{
        \compute@bb
        \if@scalefirst\if@angle
        \if@width
           \in@hundreds{\@p@swidth}{\@bbw}{\ps@bbw}
           \edef\@p@swidth{\@result}
        \fi
        \if@height
           \in@hundreds{\@p@sheight}{\@bbh}{\ps@bbh}
           \edef\@p@sheight{\@result}
        \fi
        \fi\fi
        \compute@handw
        \compute@resv}

\def\psfig#1{\vbox {
        \ps@init@parms
        \parse@ps@parms{#1}
        \compute@sizes
        \ifnum\@p@scost<\@psdraft{
                \special{ps::[begin]    \@p@swidth \space \@p@sheight \space
                                \@p@sbbllx \space \@p@sbblly \space
                                \@p@sbburx \space \@p@sbbury \space
                                startTexFig \space }
                \if@angle
                        \special {ps:: \@p@sangle \space rotate \space} 
                \fi
                \if@clip{
                        \if@verbose{
                                \ps@typeout{(clip)}
                        }\fi
                        \special{ps:: doclip \space }
                }\fi
                \if@prologfile
                    \special{ps: plotfile \@prologfileval \space } \fi
                \if@decmpr{
                        \if@verbose{
                                \ps@typeout{psfig: including \@p@sfile.Z \space }
                        }\fi
                        \special{ps: plotfile "`zcat \@p@sfile.Z" \space }
                }\else{
                        \if@verbose{
                                \ps@typeout{psfig: including \@p@sfile \space }
                        }\fi
                        \special{ps: plotfile \@p@sfile \space }
                }\fi
                \if@postlogfile
                    \special{ps: plotfile \@postlogfileval \space } \fi
                \special{ps::[end] endTexFig \space }
                \vbox to \@p@srheight sp{
                        \hbox to \@p@srwidth sp{
                                \hss
                        }
                \vss
                }
        }\else{
                \if@draftbox{           
                        \hbox{\frame{\vbox to \@p@srheight sp{
                        \vss
                        \hbox to \@p@srwidth sp{ \hss \@p@sfile \hss }
                        \vss
                        }}}
                }\else{
                        \vbox to \@p@srheight sp{
                        \vss
                        \hbox to \@p@srwidth sp{\hss}
                        \vss
                        }
                }\fi

        }\fi
}}
\psfigRestoreAt
\let\@=\LaTeXAtSign

\makeatletter%
\def\nottoobig#1{{\hbox{$\left#1\vcenter to1.111\ht\strutbox{}\right.\n@space$}}}
\makeatother%

\topsep 8pt plus2pt minus4pt   

\makeatletter 
\newcommand{\C}[1]{ {\rm {#1}} }
\newcommand{\band}{\bigwedge}
\newcommand{\Band}[3]{(\bigwedge#1\!\!:\,#2\!\!:\,#3)}
\newcommand{\bor}{\bigvee}
\newcommand{\Bor}[3]{(\bigvee#1\!\!:\,#2\!\!:\,#3)}
\newcommand{\Forall}[3]{(\forall #1\!\!:\,#2\!\!:\,#3)}
\newcommand{\Exists}[3]{(\exists #1\!\!:\,#2\!\!:\,#3)}
\newcommand{\Union}[3]{(\bigcup #1\!\!:\,#2\!\!:\,#3)}
\def\union{\,\bigcup\limits\,}
\newcommand{\true}{\mbox{\it true}}
\newcommand{\false}{\mbox{\it false}}
\newcommand{\SUM}[3]{ (\sum #1 \!\! : \, #2 \!\!:\, #3) }
\newcommand{\IFS}{\mbox{\bf if}}
\newcommand{\IF}[1]{ \mbox{\bf if} \, #1 \, \rightarrow \, }
\newcommand{\GC}[2]{ #1 \, \rightarrow \, #2 }
\newlength{\filength}
\settowidth{\filength}{\mbox{\bf f{}i}}
\newsavebox{\gcbox}
\sbox{\gcbox}{\framebox[\filength]{\rule{0ex}{2ex}}}
\newcommand{\BB}[1]{\usebox{\gcbox}\; #1 \, \rightarrow \, }
\newcommand{\FI}{\; \mbox{\bf f{}i}}
\newcommand{\Skip}{ \mbox{\bf skip} }
\newcommand{\DOS}{\mbox{\bf do}}
\newcommand{\DO}[1]{\mbox{\bf do} \, #1 \, \rightarrow \,}
\newcommand{\OD}{\mbox{\bf od}}
\newcommand{\cobegin}{{\bf cobegin}\,}
\renewcommand{\|}{\, //  \,}
\newcommand{\coend}{\,{\bf coend}}
\newcommand{\Set}[1]{ \hbox{\bf\{} #1 \hbox{\bf\}}}
\newcommand{\Bag}[1]{ \{\!| #1  |\!\}}
\newlength{\leftjustindent}
\newlength{\@leftjustindent}
\setlength{\@leftjustindent}{\leftmargin}
\def\leftjust{\let\\\@leftjustcr\let\end\@endleftjust
  \addtolength{\@leftjustindent}{\leftjustindent}
  \vcenter\bgroup
  \halign\bgroup
    \hbox to\displaywidth{
      \rule{\@leftjustindent}{0ex}$\displaystyle##$\hfill
      }\crcr
}
\def\endleftjust{\crcr\egroup\egroup\endgroup}
\def\@endleftjust#1{\crcr\egroup\egroup\@checkend{#1}\endgroup}
\def\@leftjustcr{\crcr}

\newcommand{\hoare}[3]{\{{#1}\}\:{#2}\:\{{#3}\}}
\renewcommand{\wp}[2]{ {\it wp}({#1},{#2})}
\newcommand{\assert}[1]{\!\{#1\}}
\newcommand{\atom}[1]{\langle\,{#1}\,\rangle}
\newcommand{\lbl}[1]{{#1 \!:\;\,}}
\newcommand{\pre}[1]{ {\it pre}({#1})}
\newcommand{\post}[1]{ {\it post}({#1}) }
\newcommand{\NI}[2]{ {\it NI}({#1},{#2}) }
\newcommand{\equi}[3]{ {  {\rm E}_{#1}^{#2}({#3}) }    }
\newcommand{\red}[3]{ {  {\rm R}_{#1}^{#2}({#3}) }    }
\newcommand{\sparse}{{{\rm SPARSE}}}
\newcommand{\tally}{{{\rm TALLY}}}
\newcommand{\inferfrom}[2]{\begin{array}[t]{c}\displaystyle
   \frac{#1}{#2}\end{array}}
\newtheorem{theorem}{Theorem}[section]

\newtheorem{corollary}[theorem]{Corollary}

\newcommand{\qedblob}{\mbox{\rule[-1.5pt]{5pt}{10.5pt}}}
\def\literalqed{{\ \nolinebreak\hfill\mbox{\qedblob\quad}}}
\def\qedcareful{\literalqed}
\def\qed{\literalqed}
\def\trueloveqed{{\ \nolinebreak\hfill\mbox{\boldmath
\Huge$ \Box$}\nolinebreak\mbox{$\!\!\!\!\!\!
{}^{\normalsize\heartsuit}$}}}

\newtheorem{proposition}[theorem]{Proposition}
\newcommand{\singlespacing}{\let\CS=
\@currsize\renewcommand{\baselinestretch}{1}\tiny\CS}
\newcommand{\singlespacingplus}{\let\CS=
\@currsize\renewcommand{\baselinestretch}{1.25}\tiny\CS}
\newcommand{\doublespacing}{\let\CS=
\@currsize\renewcommand{\baselinestretch}{1.75}\tiny\CS}
\newcommand{\draftspacing}{\let\CS=
\@currsize\renewcommand{\baselinestretch}{2.0}\tiny\CS}
\makeatother

\hyphenation{theory area areas theorem theorems par-allel par-allelize par-allelized threshold Hemaspaan-dra}

\newtheorem{definition}[theorem]{Definition}

\flushbottom{}
\makeatletter
\clubpenalty=\@highpenalty
\widowpenalty=\@highpenalty
\makeatother

\let\BLS=\baselinestretch
\emergencystretch=2em

\makeatletter
\newcommand{\niceonespacing}{\let\CS=\@currsize\renewcommand{\baselinestretch}{1.1}\tiny\CS}\newcommand{\nicetwospacing}{\let\CS=\@currsize\renewcommand{\baselinestretch}{1.2}\tiny\CS}
\newcommand{\nicethreespacing}{\let\CS=\@currsize\renewcommand{\baselinestretch}{1.3}\tiny\CS}
\newcommand{\singlespacingplusplus}{\let\CS=\@currsize\renewcommand{\baselinestretch}{1.35}\tiny\CS}
\newcommand{\nicefivespacing}{\let\CS=\@currsize\renewcommand{\baselinestretch}{1.5}\tiny\CS}
\newcommand{\nicesixspacing}{\let\CS=\@currsize\renewcommand{\baselinestretch}{1.6}\tiny\CS}
\newcommand{\nicefoospacing}{\let\CS=\@currsize\renewcommand{\baselinestretch}{1.15}\tiny\CS}
\newcommand{\normalspacing}{\draftspacing}
\makeatother

\singlespacing

\nicefoospacing

\makeatletter
\def\@cite#1#2{[#1\if@tempswa , #2\fi]}
\makeatother

\makeatletter
\def\@citex[#1]#2{\if@filesw\immediate\write\@auxout{\string\citation{#2}}\fi
  \def\@citea{}\@cite{\@for\@citeb:=#2\do
    {\@citea\def\@citea{,\linebreak[0]}\@ifundefined
       {b@\@citeb}{{\bf ?}\@warning
       {Citation `\@citeb' on page \thepage \space undefined}}%
\hbox{\csname b@\@citeb\endcsname}}}{#1}}
\makeatother

\newenvironment{claim}[1]{\medskip\noindent{\bf Claim #1}\it\ }{\par\medskip}
\newcommand{\claimref}[1]{#1}
\newcommand{\proofendsign}{$\rule{2mm}{2mm}$}
\newenvironment{proof}{{\noindent \bf Proof }}{{\hspace*{\fill}\proofendsign\par\bigskip}}
\newcommand{\eqq}[1]{\lq\lq #1\rq\rq}
\newcommand{\eq}[1]{\lq\lq #1\rq\rq}
\newcommand{\todopic}{{\begin{picture}(5,5)\thicklines\put(0,0){\line(1,0){5}}\put(0,0){\line(1,2){2.5}}\put(5,0){\line(-1,2){2.5}}\put(2.5,2){\makebox(0,0){{\rm !}}}\end{picture}}}
\newcommand{\todo}[1]{{\setlength{\unitlength}{1mm}\todopic}\footnote{{\setlength{\unitlength}{0.8mm}\todopic} #1}}
\newcommand{\card}[1]{||#1||}
\newcommand{\redstyle}[1]{\mathnormal{#1}}
\newcommand{\reduction}[2]{\redstyle{\le_{\mathrm{#2}}^{\mathrm{#1}}}}
\newcommand{\polyreduction}[1]{\,\reduction{p}{#1}\,}

\newcommand{\redm}{\polyreduction{m}}
\newcommand{\redctt}{\polyreduction{c}}
\newcommand{\reddtt}{\polyreduction{d}}
\newcommand{\redbtt}{\polyreduction{btt}}
\newcommand{\redbttctt}{\polyreduction{btt(c)}}
\newcommand{\redcttbtt}{\polyreduction{c(btt)}}
\newcommand{\reddttbtt}{\polyreduction{d(btt)}}
\newcommand{\redtt}{\polyreduction{tt}}
\newcommand{\redpa}{\polyreduction{\parallel}}
\newcommand{\redT}{\polyreduction{T}}
\newcommand{\redSN}{\,\reduction{SN}{T}\,}
\newcommand{\redRS}{\,\reduction{RS}{T}\,}
\newcommand{\redoverstrong}{\,\reduction{O}{T}\,}
\newcommand{\redkahorn}{\polyreduction{k{\scriptscriptstyle \!-\!}ah}}

\newcommand{\sharpp}{{\rm \#P}}
\newcommand{\sharpsat}{{\rm \#SAT}}
\newcommand{\sat}{{\rm SAT}}
\newcommand{\usatq}{{{\rm USAT}_Q}}
\newcommand{\qbf}{{\rm QBF}}
\newcommand{\parityp}{{\rm \oplus P}}
\newcommand{\up}{{\rm UP}}
\newcommand{\us}{{\rm US}}
\newcommand{\fewnp}{{\rm FewNP}}
\newcommand{\fewp}{{\rm FewP}}
\newcommand{\coup}{{\rm coUP}}
\newcommand{\e}{{\rm E}}
\renewcommand{\exp}{{\rm EXP}}
\newcommand{\NE}{{\rm NE}}
\renewcommand{\ne}{{\rm NE}}
\newcommand{\nexp}{{\rm NEXP}}
\newcommand{\p}{{\rm P}}
\newcommand{\littlep}{{\rm p}}
\newcommand{\NP}{{\rm NP}}
\newcommand{\FP}{{\rm FP}}
\newcommand{\npnp}{{\rm NP^{NP}}}
\newcommand{\bh}{{\rm BH}}
\newcommand{\BH}{{\rm BH}}
\newcommand{\BPP}{{\rm BPP}}
\newcommand{\Prob}{{\rm Prob}}
\newcommand{\MOD}{{\rm MOD}}
\newcommand{\BPTIME}{{\rm BPTIME}}
\newcommand{\ZPTIME}{{\rm ZPTIME}}
\newcommand{\DTIME}{{\rm DTIME}}
\newcommand{\dtime}{{\rm DTIME}}
\newcommand{\BPSPACE}{{\rm BPSPACE}}
\newcommand{\charfunc}[1]{\chi_{#1}}
\newcommand{\der}{\vdash} \newcommand{\notder}{\not\vdash}
\newcommand{\packalgo}{\mathtt{LearnSat}}
\newcommand{\listalgo}{\mathtt{LearnAll}}
\newcommand{\rhs}[1]{\mathrm{RHS}(#1)}

\newcommand{\ie}{{\mbox{i.e.}}}
\newcommand{\inter}{{\cap}}
\newcommand{\spp}{{\rm SPP}}
\newcommand{\gapp}{{\rm GapP}}
\newcommand{\pl}{{\rm PL}}
\def\sstar{\Sigma^{*}}
\newcommand{\R}{{\rm R}}

\newcommand{\np}{{\rm NP}}
\newcommand{\nt}{{\rm NT}}
\newcommand{\nnt}{{\rm NNT}}
\newcommand{\parityoptp}{{\rm \oplus{}OptP}}
\newcommand{\optp}{{\rm OptP}}
\newcommand{\diffp}{{\rm D^P}}
\newcommand{\pp}{{\rm PP}}
\newcommand{\bpp}{{\rm BPP}}
\newcommand{\zpp}{{\rm ZPP}}
\newcommand{\cor}{{\rm coR}}
\newcommand{\npc}{$\np$-com\-plete}
\newcommand{\conp}{{\rm coNP}}
\newcommand{\pspace}{{\rm PSPACE}}
\newcommand{\eespace}{{\rm EESPACE}}
\newcommand{\dspace}{{\rm DSPACE}}
\newcommand{\psp}{{\pspace}}
\newcommand{\pnexp}{{\p^\nexp}}
\newcommand{\npnexp}{{\np^\nexp}}
\newcommand{\nenp}{{\ne^\np}}
\newcommand{\enp}{{\e^\np}}
\newcommand{\pnp}{{\p^\np}}
\newcommand{\pnplog}{{\p^{\np[\log ]}}}
\newcommand{\pij}{{\p^{\bh_i:\bh_j}}}
\newcommand{\pji}{{\p^{\bh_j:\bh_i}}}
\newcommand{\nexpnp}{{\nexp^\np}}
\newcommand{\coNP}{{\rm coNP}}
\newcommand{\cone}{{\rm CONE}}
\newcommand{\sigmatwozero}{{\Sigma_2^0}}
\newcommand{\pitwozero}{{\Pi_2^0}}
\newcommand{\pithreezero}{{\Pi_3^0}}
\newcommand{\sigmathreezero}{{\Sigma_3^0}}
\newcommand{\sigmatwo}{{\Sigma_2^{\littlep}}}
\newcommand{\sigmathree}{{\Sigma_3^{\littlep}}}
\newcommand{\sigmafour}{{\Sigma_4^{\littlep}}}
\newcommand{\sigmafive}{{\Sigma_5^{\littlep}}}
\newcommand{\sigmak}{{\Sigma_k^{\littlep}}}
\newcommand{\sigmai}{{\Sigma_i^{\littlep}}}
\newcommand{\sigmaj}{{\Sigma_j^{\littlep}}}
\newcommand{\pitwo}{{\Pi_2^{\littlep}}}
\newcommand{\pithree}{{\Pi_3^{\littlep}}}
\newcommand{\pifour}{{\Pi_4^{\littlep}}}
\newcommand{\pifive}{{\Pi_5^{\littlep}}}
\newcommand{\thetatwo}{{\Theta_2^{\littlep}}}
\newcommand{\deltatwo}{{\Delta_2^{\littlep}}}
\newcommand{\poly}{{\rm poly}}
\newcommand{\ph}{{\rm PH}}
\newcommand{\few}{{\rm Few}}
\newcommand{\fewch}{{\rm FewCH}}
\newcommand{\eh}{{\rm EH}}
\def\bull{\vrule height .9ex width .8ex depth -.1ex }
\newcommand{\blob}{\mbox{\rule[-1.5pt]{5pt}{10.5pt}}}
\newcommand{\lindent}{\qquad}
\newcommand{\magicnum}{{ n^{\frac{1-\epsilon}{\epsilon}+\delta}}}
\newcommand{\fsup}{{\,f_{super}\,}}
\newcommand{\fred}{{\,f_{reduced}\,}}
\newcommand{\pne}{{\p^\ne}}
\newcommand{\npne}{{\np^\ne}}
\newcommand{\nnexarg}{{\nxx^\nexx (x) }}
\newcommand{\nnexx}{{\nxx^\nexx  }}
\newcommand{\nnex}{{\nxx^\nexx }}
\newcommand{\expnp}{{\exp^\np }}
\newcommand{\nxx}{{\rm N_{17}}}
\newcommand{\nexx}{{\rm NE_{21}}}
\newcommand{\seh}{{\rm SEH}}
\newcommand{\sexph}{{\rm SEXPH}}
\newcommand{\pstar}{{\p_\star}}
\newcommand{\nestar}{{\ne_{\,\star}}}
\newcommand{\supersetproper}{  \stackrel{\scriptscriptstyle\superset}{\scriptscriptstyle\not-}}
\newcommand{\subsetproper}{  \stackrel{\scriptscriptstyle\subset}{\scriptscriptstyle\not-}}
\newcommand{\superset}{\supset}
\newcommand{\superseteq}{\supseteq}

\newcommand{\substar}{\mbox{$\subset^*$}}
\newcommand{\superstar}{\mbox{$\superset^*$}}

\def\unionfromc{\,\textstyle\bigcup_{\scriptstyle c}\,}
\def\unionfromk{\,\textstyle\bigcup_{\scriptstyle k}\,}

\newcommand{\newlozenge}{\setlength{\fboxsep}{0pt}\setlength{\fboxrule}{.7pt}\framebox[6pt]{\rule{0pt}{9pt}}}

\nicefoospacing

\def\pair#1{{{\langle\!\!~#1~\!\!\rangle}}}
\def\pairs#1{{{\langle\!\!~#1~\!\!\rangle}}}
\newcommand{\piso}{\mbox{$\littlep$-iso\-mor\-phic}}
\newcommand{\manyonea}{\mbox{$\,\leq_{\rm m}^{{\littlep},\,A}\,$}}
\newcommand{\manyone}{\mbox{$\,\leq_{\rm m}^{{\littlep}}$\,}}
\newcommand{\Turing}{\mbox{$\,\leq_{\rm T}^{{\littlep}}$\,}}
\newcommand{\paiso}{\mbox{$\littlep^A$-iso\-mor\-phic}}
\newcommand{\pisoa}{\paiso}
\newcommand{\pisoam}{\mbox{$\littlep^A$-iso\-mor\-phism}}
\newcommand{\pisom}{\mbox{$\littlep$-iso\-mor\-phism}}
\newcommand{\pselective}{\mbox{$\p$-selec\-tive}}
\newcommand{\sigmastar}{\mbox{$\Sigma^\ast$}}
\newcommand{\pisnp}{\mbox{$\p=\np$}}
\newcommand{\usuba}{\mbox{$U_A$}}
\newcommand{\univsuba}{\mbox{$Univ_A$}}
\newcommand{\pisnotnp}{\mbox{$\p\neq\np$}}
\newcommand{\lb}{\mbox{\{}}
\newcommand{\rb}{\mbox{\}}}
\newcommand{\pa}{\mbox{$\p^A$}}
\newcommand{\calf}{\mbox{$\cal F$}}
\newcommand{\calc}{\mbox{$\cal C$}}
\newcommand{\cald}{\mbox{$\cal D$}}
\newcommand{\calcone}{{\cal C}_1}
\newcommand{\calctwo}{{\cal C}_2}
\newcommand{\npa}{\mbox{$\np^A$}}
\newcommand{\conpa}{\mbox{$\conp^A$}}
\newcommand{\upa}{\mbox{$\up^A$}}
\newcommand{\sparses}{\mbox{ sparse $S\,$}}
\newcommand{\bigo}{\mbox{$\cal O$}}
\newcommand{\condition}{\,\nottoobig{|}\:}
\def\land{{\; \wedge \;}}
\newcommand{\parallelnp}{\mbox{$\p_{||}^{\np}$}}
\newcommand{\rp}{\rm R}
\newcommand{\corp}{{\rm coR}}
\newcommand{\ceqp}{{\rm C_{\!=\!}P }}
\newcommand{\pclose}{\rm P-close}
\newcommand{\apt}{\rm APT}
\newcommand{\ppoly}{\rm P/poly}
\newcommand{\dr}{\mbox{\tt Carroll Ranking}}
\newcommand{\dw}{\mbox{\tt Carroll Winner}}
\newcommand{\ds}{\mbox{\tt Carroll Score}}
\newcommand{\mee}{\mbox{\tt MEE}}
\sloppy

\def\nats{\naturalnumber}
\newcommand{\naturalnumber}{\ensuremath{{  \mathbb{N} }}}
\def\wit#1{{\mbox{\rm{}WIT}_M(#1)}}
\newcommand{\acomp}{\mbox{\em acomp}}

\begin{document}

\pagestyle{plain}

\title{A Moment of Perfect Clarity II:
Consequences of Sparse Sets Hard for NP with Respect to 
Weak~Reductions\thanks{\protect\singlespacing
\copyright\ 
Christian Gla{\ss}er and
Lane A. Hemaspaandra, 
2000.
Supported in part 
by grants 
NSF-CCR-9322513 and 
NSF-INT-9815095/\protect\linebreak[0]DAAD-315-PPP-g{\"u}-ab,
and the \mbox{Studienstiftung} des \mbox{Deutschen} \mbox{Volkes}.
Written in part while Lane A.~Hemaspaandra was
visiting 
Julius-Maximilians-Universit\"at W\"{u}rzburg.}}

\author{
Christian Gla{\ss}er\thanks{\protect\singlespacing
E-mail: {\tt glasser@informatik.uni-wuerzburg.de}.}
\\  Institut f\"ur Informatik
\\  Julius-Maximilians-Universit\"at W\"{u}rzburg
\\ 97074 W\"{u}rzburg
 Germany
\and
Lane A. Hemaspaandra\footnote{\protect\singlespacing
E-mail: {\tt lane@cs.rochester.edu}.}
\\Department of Computer Science
\\University of Rochester
\\Rochester, NY 14627
USA}

\date{November 16, 2000 \protect\\ \hspace*{1.0in}}

{\singlespacing 

\maketitle

}

\begin{abstract}
   This paper discusses advances, due
    to the work of Cai, Naik, and 
Sivakumar~\cite{cai-nai-siv:tCarefulMostlyOutdatedBySTACSbutNpDttSparseImpliesUSATQinPisImplicitOnlyHereSeeComments:sparse-hard}
    and Gla{\ss}er~\cite{gla:t:sparse},
    in the complexity class
    collapses that follow if $\NP$ has sparse hard sets under
    reductions weaker than (full) truth-table reductions.
\end{abstract}


\section{Quick Hits} \label{subsec_quick_hits}

Most of this article will be devoted to presenting the work of
Gla{\ss}er~\cite{gla:t:sparse}. However, even before presenting the
background and definitions for that, let us briefly note some
improvements that follow from the
work of Cai,
Naik, and Sivakumar 
due to the 
results discussed in 
the first part of this 
article~\cite{gla-hem:j:clarityI}.
(See~\cite{gla-hem:j:clarityI}
for definitions
of the terms and classes used here:  ${\rm USAT}_Q$, 
FewP, Few, etc.)

\begin{theorem} \label{thm_sat_disjunctively_sparse_usatq_p}
    (follows from the techniques 
of~\cite{cai-nai-siv:tCarefulMostlyOutdatedBySTACSbutNpDttSparseImpliesUSATQinPisImplicitOnlyHereSeeComments:sparse-hard},
    as noted 
by~\cite{mel:perscomm:97summer,buh-for-tor:c:SixHypothesis,siv:perscomm:000506})~~If
 $\sat$ disjunctively reduces to 
a sparse set, then $(\exists Q)[\usatq \in \p]$.
\end{theorem}

A proof of
Theorem~\ref{thm_sat_disjunctively_sparse_usatq_p} 
is sketched in Section~\ref{section_proof_cainaisiv} below.
This advance of Cai, Naik, and
Sivakumar establishes immediately the following corollary
in the light of two results discussed in the first part of this 
article
(\cite{gla-hem:j:clarityI}, 
    see there for a discussion of attribution of the
first of these results), namely, 
\begin{enumerate}
\item If 
    $(\exists Q)[\usatq \in \p]$ 
    then $\p = \few$ (and thus $\p =
    \up$ and $\p = \fewp$).
\item
    \cite{val-vaz:j:np-unique}
    If $(\exists Q)[\usatq \in \p]$ then $\R = \np$.
\end{enumerate}

\begin{corollary} \label{sat_disjunctive_few}
    If $\sat$ disjunctively reduces to a sparse set, then $\p = \few$
    and $\R = \np$.
\end{corollary}

Furthermore, Arvind, K\"obler, and 
Mundhenk~\cite{arv-koe-mun:j:SparseTally} 
prove that if $\sat$ disjunctively
reduces to a sparse set, then $\ph = \p^{\np}$. However, in light of
Corollary~\ref{sat_disjunctive_few}, clearly the following can be
claimed.

\begin{theorem}
    If $\sat$ disjunctively reduces to a sparse set, then $\ph =
    \p^{\R}$.
\end{theorem}


\section{Background and Motivation}

The study of the consequences of $\np$ having sparse hard sets under
various types of (polynomial-time) reductions makes one of the most
interesting tales in complexity theory. However, we will not repeat
that tale here, as many good surveys of (parts of) that story are
available~\cite{mah:b:sparse,you:j:sparse,hem-ogi-wat:c:sparse,cai-ogi:b:sparse-hard-survey}. 
Instead,
let us cut right to the chase.

\begin{table}[t!]
    \begin{center} \renewcommand{\arraystretch}{1.25}
        \begin{tabular}{|l|l|l|}
            \hline  
            Reduction & Consequence of the existence & Reference 
\\ & of sparse sets hard for $\np$ & \\
            \hline 
            $\redm$ & $\p = \np$ & \cite{mah:j:sparse-complete} \\
            $\redbtt$ & $\p = \np$ & \cite{ogi-wat:j:pbt} \\
            $\redctt$ & $\p = \np$ & \cite{arv-han-hem-koe-loz-mun-ogi-sch-sil-thi:b:sparse}\\
            $\redbttctt$ & $\p = \np$ & \cite{arv-han-hem-koe-loz-mun-ogi-sch-sil-thi:b:sparse}\\
            $\reddtt$ & $\p^{\R} = \ph$, $\p = \few$, and $\R = \np$ & See Section~\protect\ref{subsec_quick_hits} \\
            $\redtt$ & $\ph = \zpp^{\np}$ & \cite{kob-wat:j:small-circuits} \\
            $\redT$ & $\ph = \zpp^{\np}$ & \cite{kob-wat:j:small-circuits} 
\\
            $\redRS$ & $\ph = \zpp^{\np}$ & \cite{kob-wat:j:small-circuits}, see \cite{cai-hem-wec:j:robust-reductions} \\
            $\redoverstrong$ & $\ph = \npnp$ & \cite{cai-hem-wec:j:robust-reductions} \\
            $\redSN$ & $\ph = \zpp^{\npnp}$ & Implicit in~\cite{kob-wat:j:small-circuits}, see~\cite{cai-hem-wec:j:robust-reductions} \\
\hline
        \end{tabular}
        \end{center} \caption{Consequences of the existence of sparse sets hard for $\np$.  $\redSN$, $\redoverstrong$, and $\redRS$ are 
respectively strong nondeterministic reductions;  strong and robustly
overproductive reductions; and robustly strong reductions 
(see~\protect\cite{cai-hem-wec:j:robust-reductions} for 
definitions and discussion).\label{known_results}}
\end{table}

In particular, Table~\ref{known_results} shows, for the most widely studied
reductions, the strongest currently known consequences of $\np$ having
sparse hard sets with respect to that reduction (we use the definitions
of \cite{lad-lyn-sel:j:com}, and we will, below,
define some additional reductions).

Table~\ref{known_results} brings immediately to mind the key issue: For
those reduction types for which a $\p = \np$ conclusion is not yet
known, can one achieve such a conclusion or, failing that, what is the
strongest conclusion that one can achieve?

Proving a $\p = \np$ (or even a collapse of the boolean hierarchy)
result for $\redtt$ or $\redT$ reductions may be difficult, or at
least it will require nonrelativizable techniques, due to the
following results.

\begin{theorem} \label{ten_person_paper}
    \cite{arv-han-hem-koe-loz-mun-ogi-sch-sil-thi:b:sparse}
    There is an oracle world in which $\np$ has a $\redtt$-complete
    tally set yet the boolean hierarchy does not collapse.
\end{theorem}

\begin{theorem}
    \cite{kad:j:pnplog}
    For any $f(n) = \omega(\log n)$ there is an oracle world
    in which $\np$ has sparse Turing-complete sets yet
    $\ph \neq \p^{\np[f(n)]}$.
\end{theorem}
Regarding Theorem~\ref{ten_person_paper}, one should keep in mind that
it is well-known that the following are all equivalent:
\begin{enumerate}
\item $\np$ has tally $\redtt$-hard sets.
\item $\np$
has tally $\redT$-hard sets.
\item $\np$ has sparse $\redtt$-hard sets.
\item $\np$ has sparse $\redT$-hard sets.
\end{enumerate}

Nevertheless, the gap between $\zpp^{\np}$ and smaller classes
($\p^{\np}$, $\p^{\R}$, $\np$) seems a wide one, and suggests the
importance of carefully investigating whether broad classes of
formulas formerly having only a $\zpp^{\np}$ consequence (via the
$\redtt$ line of Table~\ref{known_results}) can be shown to have
stronger consequences. Gla{\ss}er~\cite{gla:t:sparse}
has achieved
exactly this, and Section~\ref{subsec_glasser_results} will present
the key ideas of his work.


\section{Definitions} \label{subsec_definitions}

For an arbitrary set $A$ we denote the characteristic function
of $A$ by $\charfunc{A}$
and the cardinality of $A$ 
by $\card{A}$. We fix the alphabet $\Sigma = \{0,1\}$.
We denote the set of all words over $\Sigma$ 
by $\sstar$, and we denote the length of a word
$w$ by $|w|$.   We usually use \emph{language} to refer to  
(possibly nonproper) subsets of $\sstar$.
We call a set $S \subseteq \sstar$ \emph{sparse}
if and only if there
exists a polynomial $p$ such that, for all $n \ge 0$, it holds
that
$S$ contains at most $p(n)$ words whose length is no greater
than~$n$. For any sets $S_1, \ldots , S_k \subseteq \sstar$ we call
the Cartesian product
$S = S_1 \times \cdots \times S_k$ \emph{sparse} 
if and only if there exists a
polynomial $p$ such that, for all $n \ge 0$, 
it holds that $S$ contains at most $p(n)$
elements $(w_1, \ldots, w_k)$
that satisfy 
$\max\{|w_1|, \ldots, |w_k|\} \le n$.
When dealing with machines
we always talk about the deterministic version unless
nondeterminism is stated
explicitly.
We call an algorithm a \emph{$\deltatwo$~algorithm}  if it works
in polynomial time and if it has access to a $\sat$ oracle.

In boolean formulas, $\overline{v}$ denotes the negation of the
variable $v$. An \emph{anti-Horn
formula} is a boolean formula 
in conjunctive normal form such that each conjunct contains
at most one negative literal. 
We will  be particularly concerned with \emph{$k$-anti-Horn formulas};
these are, by 
definition, 
anti-Horn formulas having exactly one negative literal and at most $k$
positive literals in each conjunct. A conjunct 
$\alpha = (\overline{v_0} \vee v_1 \vee v_2 \vee \cdots \vee v_m)$
of some $k$-anti-Horn formula
is called a \emph{$k$-anti-Horn clause} and can be written as
$\alpha = (v_0 \rightarrow (v_1 \vee v_2 \vee \cdots \vee v_m))$.
We will always assume that $v_1, \ldots, v_m$ are pairwise distinct.
We refer to $v_0$ as the left-hand side of $\alpha$
and to $\{v_1, \ldots, v_m\}$
as the right-hand side of $\alpha$ ($\rhs{\alpha}$ for short). Note that
we allow empty right-hand sides, i.e., $k$-anti-Horn clauses of the form
$(v_0 \rightarrow)$;  this is  equivalent to $(\overline{v_0})$.
We write a $k$-anti-Horn formula as the set of its clauses.

We use the definitions and notations of Ladner, Lynch, and 
Selman~\cite{lad-lyn-sel:j:com} for polynomial-time reductions.
However, in case of $\redbtt$ and $\redctt$ we use the following
alternative definitions which are equivalent to those
in \cite{lad-lyn-sel:j:com}.
(For notational simplicity, henceforward
whenever we write reduction we will mean
polynomial-time reduction.)

\begin{definition}
    Let $A, B \subseteq \sstar$ be arbitrary languages.
    \begin{enumerate}
        \item \emph{$A$ 
bounded truth-table reduces to $B$}
            (denoted $A \redbtt B$) if and only if there exists
            a constant $k \ge 1$ and
            a polynomial-time machine that, given an arbitrary word $x$,
            computes a list of words $y_1, \ldots, y_k$
            and a $k$-ary boolean formula $\Phi_x$ in 
            conjunctive normal form such that
            each conjunct contains
            at most $k$ literals, and
            $x \in A \Longleftrightarrow 
            \Phi_x(\charfunc{B}(y_1), \ldots, \charfunc{B}(y_k))$.

        \item \emph{$A$ conjunctive truth-table reduces to $B$}
            (denoted $A \redctt B$) if and only if there exists a
            polynomial-time machine that, given an arbitrary word $x$,
            computes a (possibly empty, i.e.,~$m=0$) collection
            $Y_x=\{y_1, \ldots, y_m\}$ such that
            $x \in A \Longleftrightarrow (\forall j : 1\leq j \leq m)
            [\charfunc{B}(y_i)]$.
    \end{enumerate}
\end{definition}

\noindent
In addition, we define the following.
\begin{definition}
    Let $A, B\subseteq \sstar$ be arbitrary languages.
    \begin{enumerate}
        \item Let $k \ge 1$. We say that \emph{$A$ 
            $k$-anti-Horn reduces to $B$} (denoted 
          $A \redkahorn B$) if and only if
            there exists a polynomial-time machine 
            that, given an arbitrary word $x$,
            computes a list of words $y_1, \ldots, y_n$ and an $n$-ary
            $k$-anti-Horn formula $\Phi_x$ such that
            $x \in A \Longleftrightarrow \Phi_x(\charfunc{B}(y_1), \ldots, \charfunc{B}(y_n))$.
        \item $A \redbttctt B$ if and only if there exists a language $X$
            such that $A \redbtt X$ and $X \redctt B$.
        \item $A \redcttbtt B$ if and only if there exists a language $X$
            such that $A \redctt X$ and $X \redbtt B$.
        \item $A \reddttbtt B$ if and only if there exists a language $X$
            such that $A \reddtt X$ and $X \redbtt B$.
    \end{enumerate}
\end{definition}

\begin{proposition} \label{propo_cttbtt_booleanformula}
    Let $A, B \subseteq \sstar$ be arbitrary languages. 
    $A \redcttbtt B$ if and only if
    there exist a constant $k \ge 1$ and a polynomial-time machine that,
    given an arbitrary word $x$, computes a list of words $y_1, \ldots, y_n$
    and an $n$-ary boolean formula $\Phi_x$ in conjunctive normal form
    such that each conjunct contains
    at most $k$ literals, and
    $x \in A \Longleftrightarrow \Phi_x(\charfunc{B}(y_1), \ldots, \charfunc{B}(y_n))$.
\end{proposition}

We introduce the following abbreviated notation
for the case when a set $A$ reduces to a set $B$ in such a way that
for each word $x$, a list of words $y_1, \ldots, y_n$
and an $n$-ary boolean formula $\Phi_x(a_1, \ldots, a_n)$ are computed 
such that
$x \in A \Longleftrightarrow \Phi_x(\charfunc{B}(y_1), \ldots, \charfunc{B}(y_n))$.
Instead of considering the list of words $y_1, \ldots, y_n$ and the boolean formula 
$\Phi_x(a_1, \ldots, a_n)$ as separate objects,
we combine them in a natural way into a boolean formula over words, i.e.,
we replace each occurrence of some variable
$a_i$ in $\Phi_x(a_1, \ldots, a_n)$ by
the word $y_i$. For instance, if the reduction of some word $x$ produces the words
$y_1, y_2, y_3$ and the boolean formula 
$\Phi_x(a_1, a_2, a_3) = (a_1 \vee \overline{a_2}) \wedge (a_1 \vee \overline{a_3})$,
then as a simplification we assume that the reduction produces the formula
$\Phi_x = (y_1 \vee \overline{y_2}) \wedge (y_1 \vee \overline{y_3})$.
A boolean formula over words is said to be \emph{satisfied by a set $S \subseteq \sstar$}
if and only if this formula is satisfied when each occurring word $y$ is replaced
by the value $\charfunc{S}(y)$.


\section{New Collapses to $\boldsymbol{\p^{\np}}$ for Subclasses of Truth-Table Reductions} \label{subsec_glasser_results}

In this section we present the core result of~\cite{gla:t:sparse}, though 
with what we hope is a somewhat more accessible proof.
In particular, 
if there exists a sparse $\redkahorn$-hard
set for $\np$, then the polynomial hierarchy collapses to $\p^{\np}$.
From this result, the same collapse of the polynomial
hierarchy from the existence of sparse $\redcttbtt$-hard or sparse
$\reddttbtt$-hard sets for $\np$ can be shown to also hold 
(see the end of this section).

Throughout this section,
we consider only boolean formulas (respectively, boolean clauses)
over words,
$k$ will always denote the parameter of $\redkahorn$ reductions,
$p,q,r$ will denote polynomials, $v,w,x,y,z$ will denote words from
$\sstar$, $\alpha, \beta, \gamma, \delta, \theta$
will denote $k$-anti-horn clauses, $\Gamma, \Delta, \Theta$ will denote
$k$-anti-horn formulas, and ${\cal L}, {\cal L}_1, {\cal L}_2, \ldots$ will denote lists of
$k$-anti-horn formulas.

We introduce the following binary relation on $k$-anti-Horn clauses and
$k$-anti-Horn formulas.
\begin{definition}
    For $k$-anti-Horn clauses
    $\gamma = (v_0 \rightarrow (v_1 \vee \cdots \vee v_m))$ and $\delta = (w_0 \rightarrow (w_1 \vee \cdots \vee w_n))$, we write $\gamma \der \delta$ if and only if
    $v_0 \neq w_0$ or 
    $\{v_1, \ldots, v_m\} \subseteq \{w_1, \ldots, w_n\}$.
    For $k$-anti-Horn formulas $\Gamma$ and $\Delta$, we write
    $\Gamma \der \Delta$ if and only if for all $\delta \in \Delta$
    there is some $\gamma \in \Gamma$ with $\gamma \der \delta$.
\end{definition}
Note that $\der$ is reflexive, and it is even transitive if all considered
clauses have the same left-hand side. It is easy to see that
$\gamma \der \delta$ and $\Gamma \der \Delta$ are decidable in
polynomial time.

\begin{theorem} \label{khorn_theorem}
    For all $k \ge 1$, if $\sat$ $\redkahorn$ reduces to a sparse set,
    then $\ph = \p^{\np}$.
\end{theorem}
\begin{proof}
    Let $k \ge 1$ and let $S$ be a sparse set such that
    $\sat \redkahorn S$.
    Let $p$ be a polynomial such that for all $n \ge 0$
    it holds that $p(n) > 1$,
    and $S$ contains at most $p(n)$ words having length at most~$n$.
    Let $\Phi_x$ denote the $k$-anti-Horn formula that occurs when reducing
    the word $x$ to the sparse set $S$ via the $\redkahorn$
    reduction mentioned above. Moreover,
    let $q$ be a polynomial such that for all
    words $x$ it holds that (i)~$\Phi_x$ does not contain more than
    $q(|x|)$ $k$-anti-Horn clauses and (ii)~the length of words
    appearing in $\Phi_x$ is bounded by $q(|x|)$.

    Our aim is to show that $\np^{\np} = \p^{\np}$, as that
    implies that the polynomial hierarchy collapses to $\p^{\np}$.
    The proof has three parts, and in the first part we show the
    following claim.

    \begin{claim}{A}
        There exists a $\deltatwo$~algorithm $\packalgo$
        such that for all $n \in \nats$ and $z \in \sstar$ the computation $\packalgo(0^n,z)$ returns
        a $k$-anti-Horn formula $\Gamma'$ with the following properties:
        \begin{enumerate}
            \item[(i)] each clause $\gamma \in \Gamma'$ has the left-hand side $z$,
            \item[(ii)] $\Gamma'$ is satisfied by $S$, and
            \item[(iii)] $\Gamma' \der \Phi_x$ for all $x \in \sat$ with $|x| \le n$.
        \end{enumerate}
    \end{claim}

    So the output $\Gamma'$ of the computation $\packalgo(0^n,z)$ allows a
    forecast concerning queries to $\sat$ of length at most~$n$,
    in such a way that
    elements of $\sat$
    are treated correctly. Then, in the second part of the proof,
    we use Claim~\claimref{A} to show the following.

    \begin{claim}{B}
        There exists a $\deltatwo$~algorithm $\listalgo$ that, on input $0^n$, returns a list of $k$-anti-Horn
        formulas, ${\cal L}_n$, such that for all words $x \in \sstar$, $|x| \le n$, it holds that
        $x \in \sat \Longleftrightarrow (\forall \Gamma \in {\cal
        L}_n)[\Gamma \der \Phi_x]$.
    \end{claim}

    In other words, $\listalgo(0^n)$ returns a list of $k$-anti-Horn
    formulas, ${\cal L}_n$, such that each $x \in \sat$ with
    $|x| \le n$ is forecast as
    \eq{satisfiable} by all elements of ${\cal L}$, and for each
    $x \notin \sat$ with $|x| \le n$
    there is an element of ${\cal L}$ giving a negative
    forecast. So with ${\cal L}_n$ we can forecast queries to $\sat$
    of length at most~$n$,
    in such a way that {\it all} queries are treated correctly.
    Finally, in the third part of the proof, we use the algorithm
    $\listalgo$ to show that
    each language from $\np^{\np}$ can be accepted by a
    $\deltatwo$~algorithm. This implies
    $\np^{\np} = \p^{\np}$.

   \bigskip\noindent{\bf PART I:}\nopagebreak\bigskip\nopagebreak

\nopagebreak
We start with the listing of the algorithm $\packalgo$, which
    works on inputs of the form $(0^n,z)$ with
    $n \in \nats$ and $z \in \sstar$.

    \begin{enumerate}
        \item\hspace{0mm}\parbox[t]{150mm}{Algorithm: $\packalgo(0^n,z)$
            \label{epl_start}} 
        \item\hspace{0mm}\parbox[t]{150mm}{$\Gamma :=
            \{ (z \rightarrow z) \}$} \label{epl_init}
        \item\hspace{0mm}\parbox[t]{150mm}{{\tt for} $i := 0$ {\tt to}
            $\big(p(q(n))^{(k+1)}+1 \big)^{k}$} \label{loop_start_ext1a}
        \item\hspace{10mm}\parbox[t]{110mm}{{\tt if} there exists $x
            \in \sat$ with $|x| \le n$ and $\Gamma \notder \Phi_x$,
            {\tt then}
            determine the smallest such $x$, call it $\hat{x}$, {\tt else} {\tt return} $\Gamma$ and {\tt stop},
            {\tt endif}} \label{loop_start_ext1b}
        \item\hspace{10mm}\parbox[t]{110mm}{choose a $k$-anti-Horn clause $\delta
            \in \Phi_{\hat{x}}$ such that there is no $\gamma \in \Gamma$ with $\gamma
            \der \delta$} \label{epl_choose}
        \item\hspace{10mm}\parbox[t]{140mm}{ $\Gamma := (\Gamma \setminus
            \{\gamma \in \Gamma \condition \delta \der \gamma \}) \cup \{
            \delta \}$ } \label{epl_takentom}
        \item\hspace{10mm}\parbox[t]{140mm}{ {\tt if} $\card{\Gamma} =
            p(q(n))^{k+1}$ {\tt then} \label{card_cond}} 
        \item\hspace{20mm}\parbox[t]{100mm}{
            choose $\beta, \gamma \in \Gamma$ and
            a $k$-anti-Horn clause $\alpha$ such that
            $\beta \neq \gamma$, $\alpha$ is satisfied by $S$,
            $\alpha \der \beta$, $\alpha \der \gamma$, and
            $\alpha, \beta, \gamma$ have the
            left-hand side $z$} \label{packing_step}
        \item\hspace{20mm}\parbox[t]{130mm}{ $\Gamma := (\Gamma \setminus
            \{ \gamma \in \Gamma \condition \alpha \der \gamma \}) \cup \{
            \alpha \}$ } \label{replace_with_alpha}
        \item\hspace{10mm}\parbox[t]{140mm}{ {\tt endif} }
        \item\hspace{0mm}\parbox[t]{150mm}{{\tt next} $i$}
            \label{loop_end_ext1} 
        \item\hspace{0mm}\parbox[t]{150mm}{{\tt remark}
            this step will never be reached} \label{loop_end} 
    \end{enumerate}

    \begin{claim}{A1}
        Let $n \in \nats$ and $z \in \sstar$. Then, after the
        initialization of the variable $\Gamma$ in step~\ref{epl_init}, the
        following holds at the end of each step of the
        computation $\packalgo(0^n,z)$.
        \begin{enumerate}
            \item[(i)] $\Gamma$ is a set of $k$-anti-Horn clauses with
            $1 \le \card{\Gamma} \le p(q(n))^{k+1}$. \label{claima1_item1}
            \item[(ii)] All words that appear in elements of $\Gamma$
            are of length at most $\max\{q(n),|z|\}$. \label{claima1_item2}
            \item[(iii)] All clauses of $\Gamma$ have
            the left-hand side $z$.
            \label{claima1_item3}
            \item [(iv)] If $\gamma_1, \gamma_2 \in \Gamma$ and
            $\gamma_1 \der \gamma_2$ then $\gamma_1 = \gamma_2$.
        \end{enumerate}
    \end{claim}

    Right after step~\ref{epl_init} it holds that $\Gamma$ is a set
    of $k$-anti-Horn clauses.
    This is preserved by step~\ref{epl_takentom},
    since $\delta \in \Phi_x$ and $\Phi_x$ is
    a $k$-anti-Horn formula. Also step~\ref{replace_with_alpha}
    preserves this property since, by the choice of $\alpha$ and $\beta$
    in step~\ref{packing_step}, it holds that
    $\rhs{\alpha} \subseteq \rhs{\beta}$.
    Moreover,
    right after step~\ref{epl_init} we have $\card{\Gamma}=1$,
    step~\ref{epl_takentom} increases
    $\card{\Gamma}$ at most by $1$, and if $\card{\Gamma} =
    p(q(n))^{k+1}>1$ then step~\ref{replace_with_alpha} decreases
    $\card{\Gamma}$ by $1$. This shows the first statement
    of the claim, and analogously we can show the second one.
    For the third statement we note 
    that if all clauses of $\Gamma$ have the left-hand side $z$, then
    the choice of $\delta$ in step~\ref{epl_choose}
    implies that it has also the left-hand side $z$.
    Finally, using statement~(iii), we can show statement~(iv) analogously
    to the first statement.
    This proves Claim~\claimref{A1}.

    \begin{claim}{A2}
        If $\Gamma$, right before the execution of step~\ref{packing_step},
        is satisfied by
        $S$, then the choice of $\alpha$, $\beta$ and $\gamma$ in
        step~\ref{packing_step} is possible and can be carried out in
        time polynomial in $\max \{n,|z|\}$.
    \end{claim}

    So assume that we are right before the execution of
    step~\ref{packing_step}, and that $\Gamma$ is satisfied by $S$.
    For $0 \le i \le k$ let $\Gamma_i$ be the set of $k$-anti-Horn clauses
    $\gamma
    \in \Gamma$ such that there appear exactly $i$ words on the right-hand 
    side of $\gamma$.
    From Claim~\claimref{A1}(iii) it follows that $\card{\Gamma_0} \le 1$.
    Since $\sum_{i=0}^j h^i < h^{j+1}$ for all $h,j \in \nats$ with 
    $h \ge 2$, the condition
    $\card{\Gamma} = p(q(n))^{k+1}$ in step~\ref{card_cond}
    implies that there exists an
    $m>0$ such that $\card{\Gamma_m} > p(q(n))^m$. We use this fact in
    the following subprogram that shows a possible implementation of
    step~\ref{packing_step}. It assumes a read access to the program
    variables $\Gamma$ and $z$ of $\packalgo$, and it returns the required
    values $\alpha, \beta, \gamma$.
    \begin{itemize}
        \item\hspace{0mm}\parbox[t]{150mm}{$\Gamma_i := \{\gamma \in \Gamma \condition \card{\rhs{\gamma}}=i \}$ for $0 \le i \le k$}
        \item\hspace{0mm}\parbox[t]{150mm}{$j:=0$ and let $\hat{m}$ be the largest $m>0$ such that $\card{\Gamma_m} > p(q(n))^m$}
        \item\hspace{0mm}\parbox[t]{150mm}{{\tt repeat}}
        \item\hspace{10mm}\parbox[t]{140mm}{{\tt if} $j=0$ {\tt then}
        $\alpha_j := (z \rightarrow )$ {\tt else}
        $\alpha_j := (z \rightarrow (y_1 \vee y_2 \vee \cdots \vee y_j))$
        {\tt endif}}
        \item\hspace{10mm}\parbox[t]{140mm}{$j := j + 1$ and $\Delta_j := \{ \gamma \in
            \Gamma_{\hat{m}} \condition \alpha_{j-1} \der \gamma \}$}
        \item\hspace{10mm}\parbox[t]{120mm}{choose a word $y_j \notin \rhs{\alpha_{j-1}}$ that
            appears in a maximum number (note: set $n_j$ to that number) 
            of the right-hand sides of
            clauses in $\Delta_j$} 
        \item\hspace{0mm}\parbox[t]{150mm}{{\tt
            until} $n_j < \frac{\card{\Delta_j}}{p(q(n))}$}
        \item\hspace{0mm}\parbox[t]{150mm}{let $\alpha := \alpha_{j-1}$,
        choose disjoint
        $k$-anti-Horn clauses $\beta, \gamma \in \Delta_j$ and {\tt stop}.}
    \end{itemize}
    By Claim~\claimref{A1}(iii), it holds that all elements of
    $\Gamma_{\hat{m}} \subseteq \Gamma$ have the left-hand side $z$.
    Thus
    $\Delta_{j+1} = \{\gamma \in \Gamma_{\hat{m}} \condition
    y_1, y_2, \ldots, y_j 
    \mathrm{\ appear\ at\ the\ right\ hand\ side\ of\ } \gamma \}$.
    So,
    as long as the algorithm does not stop,
    it holds that $\card{\Delta_{j+1}}$ is equal to $n_j$,
    and $\card{\Delta_{j+1}} \ge \card{\Delta_{j}} / p(q(n))$.
    If we reach the $\hat{m}^{\mathrm{th}}$
    pass, we have
    $\alpha_{\hat{m}-1} = (y_1, y_2, \ldots, y_{\hat{m}-1})$ which in turn implies
    $n_{\hat{m}} = 1$
    (note that the right-hand sides of clauses in $\Gamma_{\hat{m}}$
    consist of
    exactly $\hat{m}$ elements, and we do not have two or more identical clauses since $\Gamma_{\hat{m}}$ is a set). So we obtain
    $\card{\Delta_{\hat{m}}} / p(q(n)) \ge \card{\Gamma_{\hat{m}}} / p(q(n))^{\hat{m}} > 1 =
    n_{\hat{m}}$, and it follows that the algorithm leaves the loop
    at the latest after the
    $\hat{m}^{\mathrm{th}}$ pass.
    Assume that the algorithm leaves the loop right after the
    ${j'}^{\mathrm{th}}$ pass with $1 \le j' \le \hat{m}$.
    Then we have
    $\card{\Delta_{j'}} \ge
    \card{\Gamma_{\hat{m}}} / p(q(n))^{j'-1} \ge
    \card{\Gamma_{\hat{m}}} / p(q(n))^{\hat{m}} > 1$
    (note that $\Delta_1 = \Gamma_{\hat{m}}$).
    Thus, when we have left the loop, it holds that $j=j'$, and
    there exist two disjoint
    $k$-anti-Horn clauses $\beta, \gamma \in \Delta_{j'}$.
    Since the loop's body is passed through at most $\hat{m} \le k$ 
    times, and each single step can be carried out in time
    polynomial in $\max\{n,|z|\}$ (note that by Claim~\claimref{A1}
    we have
    $\card{\Gamma} \le p(q(n))^{k+1}$ and all words appearing in
    elements of $\Gamma$ are of
    length at most~$\max\{q(n),|z|\}$), it follows that the
    above subprogram works in time
    polynomial in $\max\{n,|z|\}$. Moreover, it returns
    distinct $\beta, \gamma \in \Gamma_{\hat{m}} \subseteq \Gamma$
    and a $k$-anti-Horn
    clause $\alpha$ such that $\alpha \der \beta$, $\alpha \der
    \gamma$, and $\alpha, \beta, \gamma$ have the left-hand side $z$.

    So it remains to show that
    if the algorithm
    stops after the ${j'}^{\mathrm{th}}$ pass,
    then $\alpha_{{j'}-1} =
    (z \rightarrow (y_1 \vee y_2 \vee \cdots \vee y_{{j'}-1}))$
    is satisfied by $S$.
    Suppose that the subprogram stops after the ${j'}^{\mathrm{th}}$ pass,
    and that $\alpha_{{j'}-1}$
    is {\it not} satisfied by $S$, i.e., $z \in S$ and $y_1, \ldots,
    y_{{j'}-1} \notin S$.
    We know that all clauses of $\Gamma$ have the left-hand side $z$,
    and by assumption, $\Gamma$ is satisfied by $S$. It
    follows that each $\gamma \in \Delta_{j'} \subseteq \Gamma$ contains
    a word from $S$ on its right-hand side (remember that these words
    are no longer than $q(n)$). There are at most $p(q(n))$ words in
    $S$ that are no longer than $q(n)$. By a pigeon-hole argument
    there exists at least one word $y'_{j'} \in S$ such that $|y'_{j'}| \le q(n)$
    and $y'_{j'}$ appears in the right-hand side of at least
    $\card{\Delta_{j'}} / p(q(n))$ elements of $\Delta_{j'}$.
    From $y_1, \ldots, y_{{j'}-1} \notin S$ it follows that $y'_{j'} \notin \rhs{\alpha_{{j'}-1}}$ and
    $n_{j'} \ge \card{\Delta_{j'}} / p(q(n))$ which is a contradiction
    to our assumption that the algorithm stops after the
    ${j'}^{\mathrm{th}}$ pass. This proves Claim~\claimref{A2}.

    Using a $\sat$ oracle in combination with binary search,
    step~\ref{loop_start_ext1b} of $\packalgo$ can be carried out in
    time polynomial in $\max \{n,|z|\}$ (note that the size of $\Gamma$
    is polynomially bounded by Claim~\claimref{A1}).
    By Claim~\claimref{A2},
    also step~\ref{packing_step} can be carried out in time polynomial in
    $\max \{n,|z|\}$. This shows the first part of
    Claim~\claimref{A}, i.e., that $\packalgo$ is a
    $\deltatwo$~algorithm. The remaining part is shown in the
    following claim.

    \begin{claim}{A3}
        $\packalgo(0^n,z)$ returns
        a $k$-anti-Horn formula $\Gamma'$ such that
        \begin{enumerate}
            \item[(i)] each $\gamma \in \Gamma'$ has the left-hand side $z$,
            \item[(ii)] $\Gamma'$ is satisfied by $S$, and
            \item[(iii)] $\Gamma' \der \Phi_x$ for all $x \in \sat$ with $|x| \le n$.
        \end{enumerate}
    \end{claim}

    Assume for the moment that $\packalgo(0^n,z)$ returns some
    $\Gamma'$, i.e., $\packalgo(0^n,z)$ stops in
    step~\ref{loop_start_ext1b}.
    From Claim~\claimref{A1} it follows that statement~(i) holds,
    and that $\Gamma'$ is a $k$-anti-Horn formula.
    Clearly, $\Gamma$ is
    satisfied by $S$ after its initialization in
    step~\ref{epl_init}, and step~\ref{epl_takentom} preserves this
    property, since $\delta \in \Phi_{\hat{x}}$ and $\hat{x} \in \sat$. From
    the choice of $\alpha$ in step~\ref{packing_step}, it follows that 
    step~\ref{replace_with_alpha} also preserves the property that
    $\Gamma$ is satisfied by $S$. This shows
    (ii). Since the algorithm stops in
    step~\ref{loop_start_ext1b}, we have
    $\Gamma' \der \Phi_x$ for all $x \in \sat$ with $|x| \le n$.
    This shows (iii).

    So it remains to show that $\packalgo(0^n,z)$ stops in
    step~\ref{loop_start_ext1b}.
    Let us assign a weight $w(\theta)$ to each $k$-anti-Horn clause
    $\theta$,
    $\theta = (v_0 \rightarrow (v_1 \vee v_2 \vee \cdots \vee v_j))$,
    such that $w(\theta)$ is greater than
    the sum of the weights of $p(q(n))^{k+1}$
    (i.e., the number bounding $\card{\Gamma}$ in $\packalgo(0^n,z)$)
    $k$-anti-Horn
    clauses that have more than $j$ words on their right-hand side.
    For a $k$-anti-Horn clause $\theta$ and
    a $k$-anti-Horn formula $\Theta$ we define
    $w(\theta) = \big(p(q(n))^{(k+1)}+1 \big)^{(k-\card{\rhs{\theta}})}$
    and $w(\Theta) = \sum_{\theta \in \Theta} w(\theta)$.

    By Claim~\claimref{A1}(iii), in each step of the computation
    $\packalgo(0^n,z)$
    it holds that all clauses of $\Gamma$ have the left-hand
    side $z$. From the fact that $\delta$ (in step~\ref{epl_takentom})
    and $\alpha$ (in step~\ref{replace_with_alpha}) have
    the left-hand side $z$,
    it follows that
    $\{ \gamma \in \Gamma \condition \delta \der \gamma \} = 
    \{ \gamma \in \Gamma \condition \rhs{\delta} \subseteq \rhs{\gamma} \}$
    (in step~\ref{epl_takentom}) and
    $\{ \gamma \in \Gamma \condition \alpha \der \gamma \} = 
    \{ \gamma \in \Gamma \condition \rhs{\alpha} \subseteq \rhs{\gamma} \}$
    (in step~\ref{replace_with_alpha}).
    From the choice of $\delta$ (respectively, $\alpha$)
    it follows that it was not in $\Gamma$,
    right before step~\ref{epl_takentom}
    (respectively, step~\ref{replace_with_alpha}).
    For $\delta$ this holds by its definition in step~\ref{epl_choose},
    and for $\alpha$
    this is due to its definition in
    step~\ref{packing_step} and Claim~\claimref{A1}(iv).
    Thus, in
    step~\ref{epl_takentom} (respectively, step~\ref{replace_with_alpha})
    we add one new clause $\delta$ (respectively, $\alpha$) to $\Gamma$,
    and simultaneously we delete all clauses $\gamma \in \Gamma$ such that
    $\rhs{\delta} \subsetneq \rhs{\gamma}$ (respectively, $\rhs{\alpha}
    \subsetneq \rhs{\gamma}$).
    From Claim~\claimref{A1}(i) it follows that both steps increase the
    value of $w(\Gamma)$. Hence, each pass through the loop of
    $\packalgo(0^n,z)$ increases $w(\Gamma)$.
    We reach the highest possible value
    $w(\Gamma) = \big(p(q(n))^{(k+1)}+1 \big)^{k}$
    when $\Gamma = \{ (z
    \rightarrow) \}$. At least at this point $\packalgo(0^n,z)$ stops in
    step~\ref{loop_start_ext1b}. Since we start with
    $\Gamma = \{ (z \rightarrow z) \}$ and
    $w(\{ (z \rightarrow z) \}) \ge 0$,
    we actually reach the end of the body of the 
    loop at most $\big(p(q(n))^{(k+1)}+1 \big)^{k}$ times. Thus
    $\packalgo(0^n,z)$ stops in step~\ref{loop_start_ext1b}.
    This proves Claim~\claimref{A3}.

    This shows Claim~\claimref{A} and completes the first part
    of the proof of Theorem~\ref{khorn_theorem}.

   \bigskip\noindent{\bf PART II:}\nopagebreak\bigskip\nopagebreak

\nopagebreak
In this part we prove Claim~\claimref{B}, i.e., we construct
    a $\deltatwo$~algorithm $\listalgo$ and show that on input $0^n$ this
    algorithm will compute a list ${\cal L}_n$ of $k$-anti-Horn formulas
    $\Gamma_1, \Gamma_2, \ldots, \Gamma_m$ such that the following
    holds for all words $x$ of length at most~$n$,
    $$x \in \sat \quad \Longleftrightarrow \quad (\Gamma_i \der \Phi_x
    \mbox{ for all } 1 \le i \le m).$$
    We give the listing of the algorithm $\listalgo$, which
    works on inputs of the form $0^n$ with $n \in \nats$.
    \begin{enumerate}
        \item\hspace{0mm}\parbox[t]{150mm}{Algorithm:
            $\listalgo(0^n)$} \label{constructlistalgo_start}
        \item\hspace{0mm}\parbox[t]{150mm}{{\tt for} $i=1$
            {\tt to} $n$} \label{for_loop_1}
        \item\hspace{10mm}\parbox[t]{140mm}{${\cal L} := \emptyset$}
            \label{init_l_i} 
        \item\hspace{10mm}\parbox[t]{110mm}{{\tt while}
            there exists an $x \notin \sat$ with $|x| \le i$
            such that $\Gamma \der \Phi_x$
            for all $\Gamma \in {\cal L}$
            }
            \label{while_condition} 
        \item\hspace{20mm}\parbox[t]{130mm}{let $\hat{x}$ be the
            smallest $x$ that satisfies this condition} \label{search_smallest}
        \item\hspace{20mm}\parbox[t]{100mm}{for each word $v$ that appears on
            the left-hand side of some
            $\gamma \in \Phi_{\hat{x}}$,
            add $\Theta := \packalgo(0^i,v)$ to the list ${\cal L}$ } \label{add_to_list}
        \item\hspace{10mm}\parbox[t]{140mm}{{\tt endwhile}} \label{end_while}
        \item\hspace{10mm}\parbox[t]{140mm}{${\cal L}_i := {\cal L}$}
            \label{assignment_of_L_i} 
        \item\hspace{0mm}\parbox[t]{150mm}{{\tt next} $i$} 
        \item\hspace{0mm}\parbox[t]{150mm}{{\tt return} 
            ${\cal L}_n$} \label{constructlistalgo_end}
    \end{enumerate}

    We want to show that $\listalgo$ can be carried out in polynomial
    time if we are allowed to ask queries to a $\sat$ oracle. To do so,
    we first show that the number of passes through the while
    loop is bounded. Then we look into each single step of $\listalgo$ and show
    that it can be carried out in polynomial time (with $\sat$ as an
    oracle).

    \begin{claim}{B1}
        For any fixed $i$ (of step~\ref{for_loop_1}), the body of the 
        while loop (steps~\ref{while_condition}--\ref{end_while}) is
        passed through at most $p(q(i))$ times.
    \end{claim}

    If the condition in step~\ref{while_condition} is satisfied,
    then, right before step~\ref{add_to_list}, we have
    $\hat{x} \notin \sat$, $|\hat{x}| \le i$, and
    $\Gamma \der \Phi_{\hat{x}}$ for all $\Gamma \in {\cal L}$.
    Since $\hat{x} \notin \sat$ there exists some $\beta = (v_0
    \rightarrow (v_1 \vee v_2 \vee \cdots \vee v_l)) \in \Phi_{\hat{x}}$ that
    is not satisfied by $S$, i.e., $v_0 \in S$ and $v_1, v_2, \ldots,
    v_l \notin S$.

    From Claim~\claimref{A} it follows that
    (right before step~\ref{add_to_list})
    for each $\Gamma \in {\cal L}$ it holds that
    all clauses of $\Gamma$ have the same left-hand side and
    $\Gamma$ is satisfied by $S$.
    Choose an arbitrary $\Gamma \in {\cal L}$, and let $z$ be the left-hand 
    side of the elements of $\Gamma$.
    We want to show that $z \neq v_0$.
    From $\Gamma \der \Phi_{\hat{x}}$ it
    follows that there exists some $\gamma = (z
    \rightarrow (w_1 \vee w_2 \vee \cdots \vee w_m)) \in \Gamma$
    such that $\gamma \der \beta$.
    If $z = v_0$ then we have
    $\rhs{\gamma} \subseteq \rhs{\beta}$. Thus $w_1, w_2, \ldots, w_m
    \notin S$ and $z = v_0 \in S$.
    This contradicts the fact that $\gamma$ is satisfied by $S$.
    So $z \neq v_0$, and it follows that, in each execution of
    step~\ref{add_to_list}, we add to the list ${\cal L}$ at least one $\Theta$ whose elements have the left-hand side $v_0$ such that (i) $v_0$ does not appear as a left-hand side in some $\Gamma$ that was on the list ${\cal L}$ before, (ii) $|v_0| \le q(i)$, and (iii) $v_0 \in S$.
    This proves Claim~\claimref{B1} since the number of words in $S$
    of length at most~$q(i)$ is bounded by $p(q(i))$.

    \begin{claim}{B2}
        Consider the computation $\listalgo(0^n)$ for $n \ge 1$,
        and let $1 \le j \le n$.
        Then ${\cal L}_j$ (after its definition in
        step~\ref{assignment_of_L_i})
        is a list of $k$-anti-Horn
        formulas such that for all words $x$ of
        length $\le j$ it holds that
        $x \in \sat \Longleftrightarrow (\forall \Gamma \in {\cal
        L}_j)[\Gamma \der \Phi_x]$.
    \end{claim}

    If step~\ref{assignment_of_L_i} is executed for $i=j$, then
    the condition in
    step~\ref{while_condition} is false. Hence, for all $x$ of length 
    at most~$j$ it holds that 
    $(\forall \Gamma \in {\cal L}_j)[\Gamma \der \Phi_x] \Longrightarrow
    x \in \sat$. On the other hand, from Claim~\claimref{A} it follows
    that for all $x$ of length at most~$j$ we have
    $x \in \sat \Longrightarrow
    (\forall \Gamma \in {\cal L}_j)[\Gamma \der \Phi_x]$.
    This proves Claim~\claimref{B2}.

    \begin{claim}{B3}
        Using $\sat$ as an oracle,
        $\listalgo(0^n)$ can be carried out in time polynomial in $n$.
    \end{claim}

    By Claim~\claimref{B1}, it suffices to show that each
    single step of $\listalgo(0^n)$ can be carried out in time
    polynomial in $n$. First of all we have a look at
    step~\ref{add_to_list}. Since $|v| \le q(|\hat{x}|) \le q(i)$
    we obtain from Claim~\claimref{A}
    that $\packalgo(0^i,v)$ can be carried out in time polynomial in $i$
    (with $\sat$ as an oracle). It follows that the whole step~\ref{add_to_list} can be carried out in time polynomial
    in $i \le n$.  (Note:  we are at times being a bit informal regarding the
    uniformity that holds regarding our ``[is]... polynomial in''
    claims, but this is a common informality and our meaning should
    be clear.)

    Now let us see that we can test the condition in
    step~\ref{while_condition} with one query to $\sat$. For $i=1$
    this is trivial (without asking any question).
    If $i > 1$ then we have already computed the list
    ${\cal L}_{i-1}$. By Claim~\claimref{B2}, this list allows us to
    decide $x \in \sat$ in polynomial time for words $x$ of length
    at most~$i-1$ (note that by Claim~\claimref{A} and Claim~\claimref{B1},
    the size of ${\cal L}_{i-1}$ is polynomial in $i$).
    So using the fact that $\sat$ is self-reducible, ${\cal
    L}_{i-1}$ allows us to decide $x \in \sat$ in time polynomial in $i$
    for words $x$ of length at most~$i$. Let $N$ be the
    polynomial-time machine that achieves this, i.e., on input
    $(0^i,x,{\cal L}_{i-1})$ with $|x| \le i$ it decides $x \in
    \sat$. So the condition in step~\ref{while_condition} is
    equivalent to the following one:
    $$(\exists x \in \sstar : |x| \le i) [(0^i,x,{\cal L}_{i-1}) \notin
    L(N) \wedge (\forall \Gamma \in {\cal L})[\Gamma \der
    \Phi_x]].$$
    Since this is an $\np$ condition, it can be verified with one
    query to $\sat$. This shows that we can test the condition in
    step~\ref{while_condition} with one query to $\sat$. Analogously one
    shows that step~\ref{search_smallest} can be carried out in time
    polynomial in $i$ by asking queries to $\sat$ (we perform a binary search
    here). This proves Claim~\claimref{B3}.

    This completes the second part of the proof of
    Theorem~\ref{khorn_theorem}, since Claim~\claimref{B} follows
    from the Claims~\claimref{B2} and \claimref{B3}.

   \bigskip\noindent{\bf PART III:}\nopagebreak\bigskip\nopagebreak

\nopagebreak
So far we have shown that if $\sat$ reduces via a $\redkahorn$
    reduction to a sparse set $S$, then there exists a
    $\deltatwo$~algorithm $\listalgo$ such that, for all $n$, $\listalgo(0^n)$ returns a list of $k$-anti-Horn formulas
    and this list has the nice property that with its help we can answer
    queries to $\sat$ of length at most~$n$ in polynomial time. In the
    third part of this proof we exploit this property to show that
    each language from $\np^{\np}$ can be accepted by a
    $\deltatwo$~algorithm.  As is standard, 
    this implies a collapse of the polynomial
    hierarchy to $\p^{\np}$.

    Suppose we are given an arbitrary nondeterministic oracle Turing
    machine $M^{(\cdot)}$ and a polynomial $r$ bounding its
    computation time. We define a new machine $M'$ working on inputs
    of the form $(x,{\cal L})$ where $x$ is a word and ${\cal L}$ is a
    list of $k$-anti-Horn formulas
    (computed by $\listalgo$). On input $(x,{\cal L})$ the machine $M'$
    simulates the computation $M^{(\cdot)}(x)$ with the modification
    that queries $q$ are replaced by tests
    $(\forall \Gamma \in {\cal L})[\Gamma \der \Phi_q]$.
    It is easy to see that $L(M') \in \np$, since the mentioned test can be
    carried out in polynomial time.

    Now we can describe $N^{(\sat)}$, a {\it deterministic} polynomial-time
    Turing machine with $\sat$ oracle, which is such that $N^{(\sat)}$
    accepts the same language as does 
    $M^{(\sat)}$. On input $x$, the machine works as follows:
    \begin{itemize}
        \item[(i)] Determine ${\cal L} = \listalgo(0^{r(|x|)})$.
        \item[(ii)] Accept if and only if $(x,{\cal L}) \in L(M')$.
    \end{itemize}
    By Claim~\claimref{B},
    (i)~can be done in
    polynomial time with queries to $\sat$. Since $L(M') \in \np$,
    (ii)~can be verified in
    polynomial time with one query to $\sat$. This shows that
    $N^{(\sat)}$ works in deterministic polynomial time. Furthermore,
    by Claim~\claimref{B} we have $q \in \sat \Longleftrightarrow
    (\forall \Gamma \in {\cal L})[\Gamma \der \Phi_q]$ for all words $q$
    of length at most $r(|x|)$. Since $M^{(\sat)}(x)$ can only asks queries of
    length at most $r(|x|)$, the computation $M^{(\sat)}(x)$ is equivalent
    in outcome to that of $M'(x,{\cal L})$. It follows that $L(N^{(\sat)}) =
    L(M^{(\sat)})$. This shows $\p^{\np} = \np^{\np}$,
    and it follows that $\p^{\np} = \ph$.
\end{proof}

The following theorem strengthens Theorem~\ref{khorn_theorem}, i.e., it holds that
$\p^{\np} = \ph$ even if $\sat$ reduces via a $\redcttbtt$ reduction
to a sparse set. Here the reduction formulas are (unbounded)
conjunctions of formulas of bounded length.

\begin{theorem}
    If $\sat$ reduces via a $\redcttbtt$ reduction to a sparse set,
    then $\p^{\np} = \ph$.
\end{theorem}
\begin{proof}
    Suppose $\sat$ reduces via a $\redcttbtt$ reduction to a sparse
    set $S$.
    From Proposition~\ref{propo_cttbtt_booleanformula} it follows that there
    exist a constant $k \ge 1$ and a polynomial-time machine that,
    given an arbitrary word $x$, computes a boolean formula of words,
    $\Phi_x$, in conjunctive normal form such that each
    conjunct contains
    at most $k$ words, and
    $x \in \sat$ if and only if $\Phi_x$ is satisfied by $S$.
    Observe that
    there is a bijection $f : (\sstar)^0 \cup (\sstar)^1 \cup \cdots
    \cup (\sstar)^k \longrightarrow \sstar$ that is polynomial-time
    computable and polynomial-time invertible. For $0 \le i \le k$, let
    $S_i = f(S^i) = \{ f(w_1, w_2, \ldots, w_i) \condition w_1, w_2,
    \ldots, w_i \in S \}$, and note that, for each $0\leq i \leq k$, 
    $S^i$ is sparse since $S$ is sparse.
    Since $f$ is polynomial-time invertible, one can also show 
    that $S_i$ is sparse.
    Thus $S' = S_0 \cup S_1 \cup \cdots \cup S_k$ is sparse.

    We want to show that $\sat \redkahorn S'$. To do so, we consider
    the conjuncts of $\Phi_x$ for some word $x$.
    For each conjunct we perform the following transformation:
    $$(\overline{v_1} \vee \overline{v_2} \vee \cdots \vee
    \overline{v_i} \vee w_1 \vee w_2 \vee \cdots \vee w_j) \;\;\mapsto\;\;
    (\overline{f(v_1, v_2, \ldots, v_i)} \vee f(w_1) \vee f(w_2) \vee
    \cdots \vee f(w_j)).$$
    It is easy to see that this transformation can
    be carried out in polynomial time. Moreover, it holds that
    a conjunct is satisfied by $S$ if and only if the transformed conjunct
    is satisfied by $S'$
    (note that $v_1 \notin S \vee \cdots \vee v_i \notin S \Longleftrightarrow 
    f(v_1, \ldots, v_i) \notin S'$,
    and $w \in S \Longleftrightarrow f(w) \in S'$
    for all words $w, v_1, \ldots, v_i$).
    This shows that $\sat$ reduces via a
    $\redkahorn$ reduction to a sparse set $S'$. From
    Theorem~\ref{khorn_theorem} it follows that $\p^{\np} = \ph$.
\end{proof}

The proofs in this section show even more: It turns out that
we can replace $\sat$ by any polynomial-time length-decreasing
self-reducible set.
(A language $T$ is called polynomial-time length-decreasing
self-reducible if and only if 
there exists a polynomial-time oracle machine $M^{(\cdot)}$
such that the language accepted by $M^{(T)}$ is equal to $T$, and on input $x$
this machine queries the oracle only about
words $w$ with $|w| < |x|$).

\begin{theorem} \label{general_result}
    If a polynomial-time length-decreasing self-reducible set $T$ reduces via a
    $\redcttbtt$ reduction to a sparse set, then $\np^{T}
    \subseteq \p^{\np}$.
\end{theorem}

Note that if $T$ is a polynomial-time length-decreasing self-reducible set,
then also
the complement $\overline{T}$ is polynomial-time length-decreasing
self-reducible. Moreover, $T \reddttbtt S$ implies $\overline{T}
\redcttbtt S$ for any set $S$
(just by negation of the reduction formulas).
Thus for $\reddttbtt$ reductions a
result analogous to Theorem~\ref{general_result} holds.


\section{On Theorem~\protect\ref{thm_sat_disjunctively_sparse_usatq_p}}
\label{section_proof_cainaisiv}
As noted earlier, 
Theorem~\ref{thm_sat_disjunctively_sparse_usatq_p} follows from
the techniques 
of~\cite[Theorem~10]{cai-nai-siv:tCarefulMostlyOutdatedBySTACSbutNpDttSparseImpliesUSATQinPisImplicitOnlyHereSeeComments:sparse-hard},
as has been mentioned by~\cite{mel:perscomm:97summer,buh-for-tor:c:SixHypothesis,siv:perscomm:000506}.
For completeness, we note that one can 
see this, for example, as follows.
If $\sat$ disjunctively reduces to some sparse set $S$, then also the following
set $L \in \np$ disjunctively reduces to $S$ say via some function $f \in \FP$
(cf.~\cite[Theorem~10]{cai-nai-siv:tCarefulMostlyOutdatedBySTACSbutNpDttSparseImpliesUSATQinPisImplicitOnlyHereSeeComments:sparse-hard}). 
Let $$L =
 \{ \langle \psi, 1^m, u, v \rangle \;\condition\;~~~~~~~~~~~~~~~~~~~~~~~~~~~~~~~~~~~~~~~~~~~~~~~~~~~~~~~~~~~~~~~~~~~ $$
\vspace*{-0.4in}
$$ ~~~~~~~~~~~~~~~~~~~~~~~ m = 2 \cdot 3^l,\; u,v \in GF(2^m),\;
(\exists \vec{a} = (a_0, \ldots, a_{n-1}))[\psi(\vec{a}) \wedge \sum_{i=0}^{n-1} a_i u^i = v]
\},$$
where $\psi$ is an $n$-ary boolean formula, $\vec{a}$ an assignment 
for $\psi$, and $u$ and $v$ are elements
of the finite field that 
has $2^m$ elements ($m$ is of the form $2 \cdot 3^l$ for some $l \ge 0$
to guarantee that this field exists). We assume that for some given word $x$,
$f(x)$ is a set of words (that is interpreted as a disjunction of words).

Now we follow the proof of Theorem~9 which can be found in Appendix B 
of~\cite{cai-nai-siv:tCarefulMostlyOutdatedBySTACSbutNpDttSparseImpliesUSATQinPisImplicitOnlyHereSeeComments:sparse-hard}.
Let $q$ be a polynomial such that for all $n',m \ge 0$ and all
boolean formulas $\psi$ of size $n'$ it holds that the 
words in $f(\langle \psi, 1^m, u, v \rangle)$ are of length
at most~$q(n',m)$.
Moreover, let $p$ be a polynomial such that the number of strings in $S$
of length at most $q(n',m)$ is bounded by $p(n',m)$ for all $n',m \ge 0$.

Let $\phi$ be an $n$-ary boolean formula of size $n' \ge n$
that has exactly one satisfying assignment; we will
determine this assignment.
Choose the smallest suitable $m$
(i.e., $m = 2 \cdot 3^l$ for some $l \ge 0$)
such that $2^m / p(n',m) \ge n$,
call it $\hat{m}$, 
and let $F = GF(2^{\hat{m}})$. Note that $\hat{m} = O(\log n')$.

Instead of estimating probabilities as it is done in the original proof of
Theorem~9~\cite{cai-nai-siv:tCarefulMostlyOutdatedBySTACSbutNpDttSparseImpliesUSATQinPisImplicitOnlyHereSeeComments:sparse-hard} 
let us proceed as follows.
For all $u,v \in F$ we compute $f(\langle \phi, 1^{\hat{m}}, u, v \rangle)$ in polynomial time.
Since $\phi$ has exactly one satisfying assignment, for each $u \in F$
there is a unique $v_u \in F$ such that $\langle \phi, 1^{\hat{m}}, u, v_u \rangle \in L$.
For each $u$, let $S_u = \bigcup_{v \in F} f(\langle \phi, 1^{\hat{m}}, u, v \rangle)$.
Now observe the following facts.
\begin{enumerate}
    \item For each $u \in F$ there exists an $s \in S \cap S_u$
    with $|s| \le q(n',\hat{m})$.
    \item The number of elements in $S$ that are of length
    at most $q(n',\hat{m})$ is bounded by $p(n',\hat{m})$.
\end{enumerate}
It follows that there is some $w \in S$ that 
appears in at least $2^{\hat{m}} / p(n',\hat{m}) \ge n$ sets $S_u$.

For each word $w$ that 
appears in at least $n$ sets $S_{u_1}, \ldots, S_{u_n}$,
we do the following: We determine corresponding words $v_1, \ldots, v_n$,
such that $w \in f(\langle \phi, 1^{\hat{m}}, u_i, v_i \rangle)$.
Then we solve the following
equation for $\vec{a} = (a_0, a_1, \ldots, a_{n-1})$
(this is possible since we have a Vandermonde matrix).
\begin{equation} \label{eqn_cns}
    (a_0, a_1, \ldots, a_{n-1}) \cdot \left(
        \begin{array}{cccc}
            (u_1)^0 & (u_2)^0 & \cdots & (u_n)^0 \\
            (u_1)^1 & (u_2)^1 & \cdots & (u_n)^1 \\
            \vdots& \vdots&  & \vdots\\
            (u_1)^{n-1} & (u_2)^{n-1} & \cdots & (u_n)^{n-1}
        \end{array}
    \right)
    = (v_1, v_2, \ldots, v_n)
\end{equation}
Finally we check whether $\vec{a}$ is a satisfying assignment for $\phi$
and output $\vec{a}$ in this case.

Note that if we reach some $w \in S$, then all corresponding
$\langle \phi, 1^{\hat{m}}, u_i, v_i \rangle$ are elements of $L$.
By the definition of $L$ and the fact that $\phi$ has
exactly one satisfying assignment $(a_0, a_1, \ldots, a_{n-1})$,
we have $\sum_{j=0}^{n-1} a_j u_i^j = v_i$
for all $i$. So if $w \in S$, then (\ref{eqn_cns}) is a valid equation.
Thus, we really do find the satisfying assignment of $\phi$.
This shows that $(\exists Q)[\usatq \in \p]$.

\bigskip

{\bf Acknowledgments}~~The authors thank 
Matthias Galota,
Edith Hemaspaandra,
Harald Hempel,
Heinz Schmitz,
Klaus W.~Wagner,
and Gerd Wechsung
for helpful discussions and comments.


\newcommand{\etalchar}[1]{$^{#1}$}

\end{document}